\documentclass[aps,prc,twocolumn,superscriptaddress,amsfonts, amsmath,amssymb,showpacs, floatfix
]{revtex4-2}

\usepackage{graphicx}
\usepackage{dcolumn}
\usepackage{bm}
\usepackage{xcolor}
\usepackage{subcaption}
\usepackage{ulem}
\usepackage{multirow}

\def\Si{{$^{26}${\rm Si}}~}

\def\reac{{$^{22}${Mg($\alpha$,$p$)}$^{25}${Al}}~}

\def\ap{($\alpha$,$p$)~}

\def\ea{\textit{et al.}~}

\begin{document}

\title{Study of the $^{22}$Mg waiting point relevant for x-ray burst nucleosynthesis via the \boldmath{$^{22}${Mg($\alpha$,$p$)}$^{25}${Al}} reaction}

\author{H. Jayatissa}
\email{Corresponding author \newline hesh.jayatissa@gmail.com}
\affiliation{Physics Division, Argonne National Laboratory, Lemont, IL, 60439, USA}

\author{M.~L.~Avila}
\affiliation{Physics Division, Argonne National Laboratory, Lemont, IL, 60439, USA}

\author{K.~E.~Rehm}
\affiliation{Physics Division, Argonne National Laboratory, Lemont, IL, 60439, USA}

\author{P.~Mohr}
\affiliation{Institute for Nuclear Research (Atomki), P.O. Box 51, Debrecen H-4001, Hungary}

\author{Z.~Meisel}
\affiliation{Institute of Nuclear and Particle Physics, Ohio University, Athens, OH 45701, USA}

\author{J.~Chen}
\affiliation{Physics Division, Argonne National Laboratory, Lemont, IL, 60439, USA}

\author{C.~R.~Hoffman}
\affiliation{Physics Division, Argonne National Laboratory, Lemont, IL, 60439, USA}

\author{J.~Liang}
\altaffiliation[Present address: ]{TRIUMF, Vancouver, British Columbia, V6T 2A3, Canada}
\affiliation{Department of Physics \& Astronomy, McMaster University, Hamilton, Ontario L8S 4M1, Canada}

\author{C.~M\"{u}ller-Gatermann}
\affiliation{Physics Division, Argonne National Laboratory, Lemont, IL, 60439, USA}

\author{D.~Neto}
\affiliation{Department of Physics, University of Illinois Chicago, 845 W. Taylor St., Chicago, IL 60607, USA}

\author{W.~J.~Ong}
\affiliation{Lawrence Livermore National Laboratory, 7000 East Ave, Livermore, CA 94550, USA}

\author{A.~Psaltis}
\affiliation{Institut für Kernphysik, Technische Universität Darmstadt, Darmstadt D-64289, Germany}

\author{D.~Santiago-Gonzalez}
\affiliation{Physics Division, Argonne National Laboratory, Lemont, IL, 60439, USA}

\author{T.~L.~Tang}
\altaffiliation[Present address: ]{Department of Physics, Florida State University, Tallahassee, Florida 32306, USA}
\affiliation{Physics Division, Argonne National Laboratory, Lemont, IL, 60439, USA}

\author{C.~Ugalde}
\affiliation{Department of Physics, University of Illinois Chicago, 845 W. Taylor St., Chicago, IL 60607, USA}

\author{G.~Wilson}
\affiliation{Department of Physics and Astronomy, Louisianna State University, Baton Rouge, LA 70803, USA}

\begin{abstract}

The \reac reaction rate has been identified as a major source of uncertainty for understanding the nucleosynthesis flow in Type-I x-ray bursts (XRBs). We report a direct measurement of the energy- and angle-integrated cross sections of this reaction in a 3.3--6.9~MeV center-of-mass energy range using the MUlti-Sampling Ionization Chamber (MUSIC). The new \reac reaction rate is a factor of $\sim$4 higher than the previous direct measurement of this reaction within temperatures relevant for XRBs, resulting in the $^{22}$Mg waiting point of x-ray burst nucleosynthesis flow to be significantly bypassed via the ($\alpha,p$) reaction.

\end{abstract}

\maketitle

An x-ray burst is a thermonuclear explosion in a binary system of an accreting neutron star and a companion star \cite{Lewin1993,Schatz2006,Parikh2013}. Properties of the neutron star (e.g compactness, crust composition, mass-radius relation, etc.) can be deduced from comparisons between the observations of XRB light curves and astrophysical models \cite{Schatz1998,Fisker2008,Meisel2018, Meisel20182}. These models significantly depend on nuclear physics inputs, such as nuclear reaction rates \cite{Cyburt2016, Meisel2019}. The accretion of hydrogen from the companion star prior to an XRB ensures that the main nucleosynthesis occurs via proton capture on elements formed via the hot CNO cycle and its breakout reactions \cite{Wiescher1999}. It has been suggested that the main ($p,\gamma$) nucleosynthesis path in XRBs are halted at several ``waiting points" \cite{Fisker2004} due to ($p,\gamma$)-($\gamma,p$) equilibrium. Alpha capture reactions could allow the halted nucleosynthesis process to bypass these waiting points to synthesize heavier elements.

One of the waiting points identified for XRB nucleosynthesis is $^{22}$Mg (T$_{1/2}$ = 3.876 s) \cite{BASUNIA2015}. The interplay between the $(\alpha,p)$ reaction with the proton capture reaction on $^{22}$Mg and the subsequent $\beta$-decay plays an important role for the subsequent nucleosynthesis flow. Since the $Q$-value of the $^{22}$Mg($p,\gamma$)$^{23}$Al reaction is small ($Q$-value = 0.141~MeV) \cite{Huang2021}, a ($p,\gamma$)-($\gamma,p$) equilibrium is established \cite{Meisel2018}, and capture reactions will occur within timescales that are short compared to the half life of $^{22}$Mg. Hence, the reaction flow through the $^{22}$Mg waiting point is mainly determined by the $^{22}$Mg($\alpha,p$)$^{25}$Al reaction and the proton capture rate on $^{23}$Al created via $^{22}$Mg($p,\gamma$)$^{23}$Al reaction. The uncertainties on the proton capture rate on $^{23}$Al have recently been substantially reduced experimentally \cite{Puentes2022, Lotay2022}. Currently, the \reac reaction rate provides the main uncertainty in constraining the nucleosynthesis flow at the $^{22}$Mg waiting point, despite recent experimental efforts. Moreover, sensitivity studies have identified the \reac reaction to significantly impact the XRB light curves and burst ashes \cite{Cyburt2010,Cyburt2016}. 

Several experiments have been carried out to constrain the \reac reaction rate. An indirect measurement was carried out by Matic \textit{et al.} \cite{Matic2011} using the $^{28}$Si($p,t$)$^{26}$Si reaction to study the level structure of \Si. This work identified four resonances in \Si above the $\alpha$-decay threshold ($Q_{\alpha}$ = 9.166~MeV), with unknown spins and parities. Due to the limited information obtained in this work, the \reac rate was deduced with large uncertainties and it was found to be several orders of magnitude lower than the predictions using the Hauser-Feshbach (HF) formalism. Later, the first direct measurement of the \reac reaction was performed in inverse kinematics by Randhawa \textit{et al.} \cite{Randhawa2020}. This work utilizes recoil protons measured within a limited angular range and the PACE4 code to extract the angle-integrated cross sections in a center-of-mass energy range of 3.2--10.6~MeV. An uncertainty of $\sim$35\% was estimated in the cross sections due to the uncertainty from the model prediction of the proton angular distributions. In order to fit the low reaction cross sections and to extrapolate them down to the center-of-mass energies relevant for XRBs, the code TALYS \cite{TALYS} was used with notable modifications to the width of the $\alpha$-particle optical model potential ($\alpha$OMP) applied, and an increase of the default \Si level density by a factor of 2.6. When compared to the HF predictions using the code NON-SMOKER \cite{NON_SMOKER}, the total reaction cross sections obtained from this work is a factor of $\approx$8 lower. Recently, Hu \textit{et al.} \cite{Hu2021} carried out a measurement of the $^{25}$Al+$p$ (in)elastic scattering reaction in the excitation energy range between 9--11 MeV. This work performed an R-matrix fit to assign spins and parities of the states observed. The partial $\alpha$ widths of four resonances above the $\alpha$-decay threshold in \Si were inferred from the mirror nucleus $^{26}$Mg. Surprisingly, the number of measured resonances is low when compared with the almost 100 known levels in the same excitation energy range of the mirror nucleus $^{26}$Mg. Assuming that at least one half of these levels in $^{26}$Mg have unnatural parity, the number of expected candidate levels in \Si is a factor of $\sim$10 higher than the number of resonances measured by Hu \textit{et al.} \cite{Hu2021}. Thus, the rate from this work should be considered as a lower limit. The \reac reaction rate from that experiment is a factor of 8--10 lower than that obtained from the HF code NON-SMOKER for $\sim$0.4--1~GK, and increases up to a factor of $\sim$160 at 3~GK. For temperatures below 1~GK, this rate is in reasonable agreement with that by Randhawa \textit{et al.} \cite{Randhawa2020}. From the low reaction rate found by these two previous measurements, Ref. \cite{Hu2021} concluded that the XRB reaction flow follows mainly the $^{22}$Mg($p,\gamma$)$^{23}$Al($p,\gamma)^{24}$Si path, implying that the $^{22}$Mg waiting point is potentially not bypassed. Further astrophysical implications of the low \reac rate from this work can be found in Ref. \cite{Lam2022}.

In this work, we present a new independent study of the $^{22}$Mg$(\alpha,p)^{25}$Al reaction that directly measures angle- and energy-integrated cross sections, removing the model dependence for obtaining the total cross sections, common to all previous measurements. A direct measurement of the \reac reaction in inverse kinematics was performed using the Argonne Tandem Linac Accelerator System (ATLAS) at Argonne National Laboratory. An in-flight radioactive beam of $^{22}$Mg$^{12+}$ was developed with the ATLAS in-flight system \cite{RAISOR} using the $^{20}$Ne($^{3}$He,$n$)$^{22}$Mg reaction with a primary beam of $^{20}$Ne at an energy of 125.0~MeV. The $^{22}$Mg beam, with an energy of 74.0 $\pm$ 1.5~MeV, an average intensity of 200 pps, and a purity of $\approx$18 \%, was delivered to the MUlti-Sampling Ionization Chamber (MUSIC) detector filled with 404 Torr of pure He gas. More information on the MUSIC detector can be found in Ref. \cite{MUSIC_NIM}. The $^{22}$Mg beam energy was determined using a silicon detector upstream of MUSIC and confirmed by the known magnetic rigidity of an upstream beam-line transport dipole magnet. The silicon detector calibration was based on the measured energy of the unreacted $^{20}$Ne beam as determined by the ATLAS time-of-flight system. The main contaminants of the $^{22}$Mg beam detected within MUSIC include different charge states of the primary $^{20}$Ne beam and a small amount of $^{22}$Na$^{10+}$. Using the energy deposited in the Frisch grid and the first anode strip, the $^{22}$Mg beam can be separated from these contaminants, and the total amount of incoming $^{22}$Mg beam is obtained for normalization of the cross section. 

Energy loss tables are used to calculate the energy of the beam at each of the anode strips of the detector. In order to identify the best energy loss table for the present work, a silicon detector was placed downstream of MUSIC to measure the remaining $^{22}$Mg beam energy. Since the beam stops inside the detector for the He gas pressure used for this measurement, the pressures were lowered to allow the $^{22}$Mg beam to go through the gas and exit window foil of the MUSIC detector. The gas pressures used ranged from 0 to 250~Torr in steps of 50~Torr. The beam energies measured with this silicon detector were well reproduced using energy loss calculations performed using the ATIMA 1.2 \cite{atima} energy loss tables. The choice of these tables was also cross-checked via its reproducibility of the Bragg peak in the energy loss of the beam inside MUSIC for the 404 Torr gas pressure used for the final measurement. This work presents a direct \reac cross section measurement in the center-of-mass energy range of 3.3--6.9~MeV. 

\begin{figure}[!ht]
\includegraphics[width=0.5\textwidth]{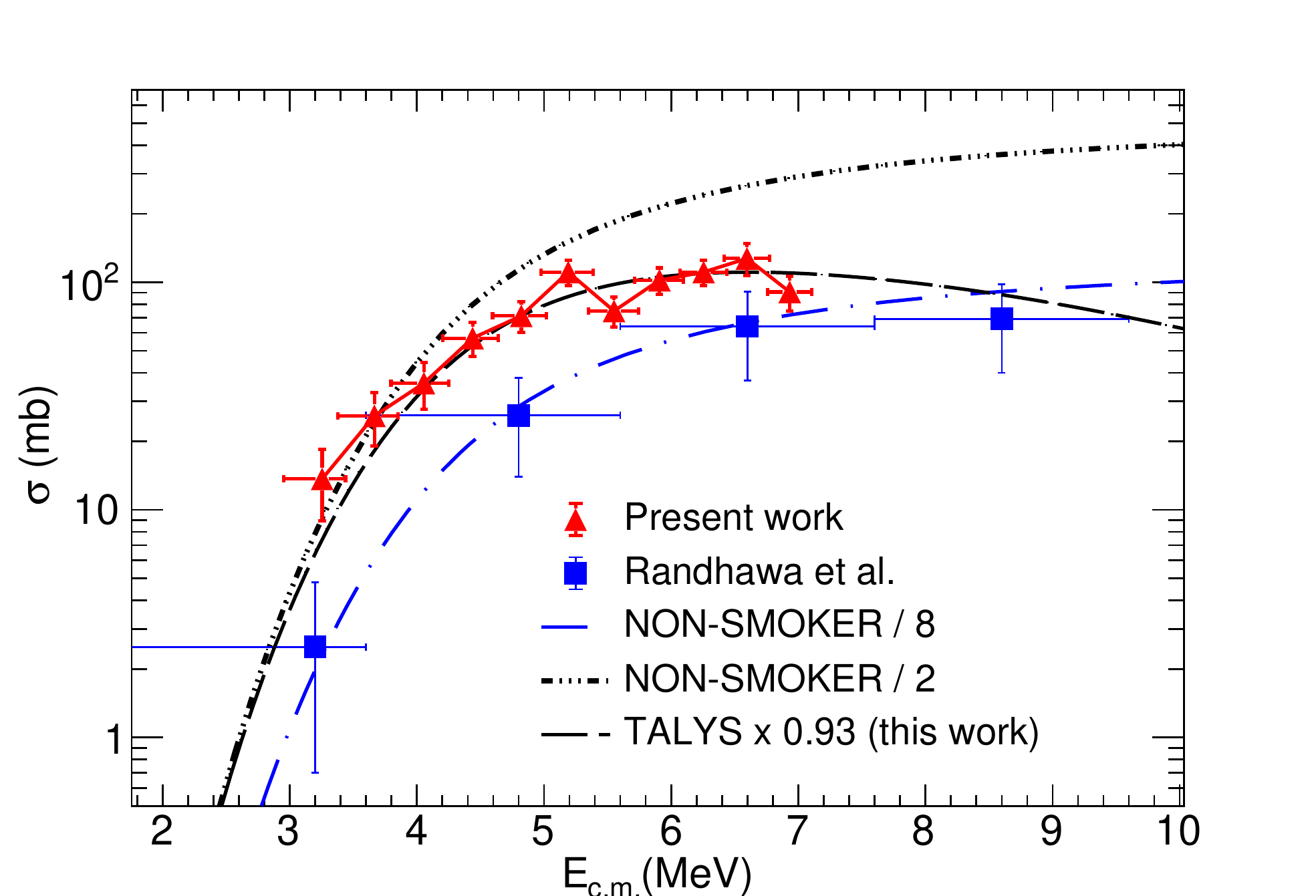}
\caption{\label{fig:CS} Experimental cross sections from the present measurement (red triangles) and the data from Ref. \cite{Randhawa2020} (blue squares). Theoretical HF cross section calculations using TALYS (scaled by 0.93), NON-SMOKER/8 and NON-SMOKER/2 are shown.}
\end{figure}

The MUSIC detector is sensitive to the energy loss of a particle as it travels through the gas and has the ability to measure an excitation function with a single beam energy. When a reaction occurs at any anode strip, differences in the energy deposited in each strip can be used to separate out the $(\alpha,p)$ events of interest from the other reaction channels. By summing the energy deposited in various numbers of consecutive strips using a $\Delta E-\Delta E$ technique after a reaction occurs, allows for further separation of the different reaction channels (see Ref. \cite{Jayatissa2022}). More information on using MUSIC for direct $\alpha$-induced measurements can be found in Refs.~\cite{Avila2017,Jayatissa2022,Ong2022,Talwar2018,Avila2016}. For the energy range covered in this work, the $(\alpha,\gamma)$, $(\alpha,p)$, $(\alpha,2p)$, elastic $(\alpha,\alpha)$ and inelastic $(\alpha,\alpha')$ channels are energetically allowed. The MUSIC detector is not sensitive to the $(\alpha,2p)$ reaction channel, since the energy losses in the detector gas from this channel cannot be separated from the $(\alpha,\alpha)$ or $(\alpha,\alpha')$ channels.

\begin{table}[ht]
\caption{\label{tab:CS} Total reaction cross sections, $\sigma$, and associated systematic and statistical uncertainties obtained from the present measurement for the \reac reaction for center-of-mass energies corresponding to anode strips 2-12 of MUSIC. }
\begin{ruledtabular}
\begin{tabular}{ccccc}
\begin{tabular}[c]{@{}c@{}}E$_{\textrm{c.m.}}^{\textrm{eff}}$\\ (MeV)\end{tabular} &
  \begin{tabular}[c]{@{}c@{}}$\Delta$E$_{\textrm{c.m.}}$\footnote{The energy binning per strip is determined by the energy loss of the $^{22}$Mg beam along the width of each corresponding strip.}\\ (MeV)\end{tabular} &
  \begin{tabular}[c]{@{}c@{}}$\sigma$\\ (mb)\end{tabular} &
  \begin{tabular}[c]{@{}c@{}}$\Delta\sigma_{sys}$\\ (mb)\end{tabular} &
  \begin{tabular}[c]{@{}c@{}}$\Delta\sigma_{stat}$\\ (mb)\end{tabular} \\
  \hline \\\vspace{2mm}
6.93 (28) & $\substack{+0.17 \\ -0.17}$ & 91 & 11 & 11 \\
\vspace{2mm} 6.60 (29) & $\substack{+0.18 \\ -0.18}$ & 127 & 15  & 14 \\ 
\vspace{2mm} 6.25 (29) & $\substack{+0.18 \\ -0.18}$ & 111 & 7  & 13 \\ 
\vspace{2mm} 5.91 (30) & $\substack{+0.19 \\ -0.19}$ & 102 & 6  & 12 \\ 
\vspace{2mm} 5.55 (31) & $\substack{+0.19 \\ -0.20}$ & 75 & 5  & 10 \\ 
\vspace{2mm} 5.19 (31) & $\substack{+0.19 \\ -0.21}$ & 111 & 7  & 13 \\ 
\vspace{2mm} 4.82 (32) & $\substack{+0.20 \\ -0.23}$ & 71 & 4  & 10 \\ 
\vspace{2mm} 4.44 (33) & $\substack{+0.20 \\ -0.24}$ & 57 & 3  & 9 \\ 
\vspace{2mm} 4.06 (33) & $\substack{+0.19 \\ -0.26}$ & 36 & 4 & 7 \\ 
\vspace{2mm} 3.67 (34) & $\substack{+0.18 \\ -0.29}$ & 26 & 3 & 6  \\
\vspace{2mm} 3.25 (35) & $\substack{+0.18 \\ -0.30}$ & 14 & 2 & 4  \\
\end{tabular}
\end{ruledtabular}
\end{table}

For the center-of-mass energies corresponding to each strip, the total \ap cross section can be obtained by normalizing the total number of \ap events identified for each strip to the measured beam intensity. Figure \ref{fig:CS} shows the total reaction cross sections obtained from the present work in comparison with the direct measurement of Ref. \cite{Randhawa2020} and scaled theoretical HF calculations using TALYS and NON-SMOKER. The corresponding statistical and systematic uncertainties of the cross sections from the present work are tabulated in Table \ref{tab:CS}. Here, the effective center-of-mass energies ($E_{\textrm{c.m.}}^{\textrm{eff}}$) have been adjusted to take into account the energy dependence of the cross sections over the target thickness for each data point. The uncertainties of $E_{\textrm{c.m.}}^{\textrm{eff}}$ is defined by the uncertainty of the beam energy. The estimated energy loss of the beam within each anode strip defines the energy binning $\Delta E_{\textrm{c.m.}}$. The systematic uncertainty of the cross sections arises predominantly from the analysis techniques and conditions used to separate the \ap events from the beam and other reaction channels. This separation becomes more difficult at the beginning of the detector and as the reaction vertex gets closer to the Bragg peak of the beam. Thus higher systematic uncertainties are assigned for the first and last two energy points.

As seen in Fig. \ref{fig:CS}, the NON-SMOKER cross sections have been divided by a factor of 8 to reproduce the data of Ref. \cite{Randhawa2020}. However, it is important to point out that the NON-SMOKER calculations do not differentiate between the contributions from the $(\alpha,p)$ and $(\alpha,2p)$ channels, which do not represent these experimental data. The total reaction cross section of $^{22}$Mg + $\alpha$ calculated using TALYS depends only on the chosen $\alpha$OMP. The $\alpha$OMP by McFadden and Satchler \cite{MCFADDEN1966} was used in the present study since this potential reproduces the experimental data for masses $20 \lesssim A \lesssim 50$ with typical uncertainties below a factor of two \cite{Mohr2015}. The statistical model distributes the total cross section $\sigma$ among the open reaction channels in the energy range under study. The contribution of the ($\alpha$,$\gamma$) channel to the cross section remains very minor (smaller by $\approx$ 5 orders of magnitude). The theoretical $^{25}$Al production cross section in the $^{22}$Mg($\alpha,p$)$^{25}$Al reaction, as measured by MUSIC, was calculated by the sum over the final states in $^{25}$Al; $\sigma(^{25}{\rm{Al}}) = \sum_i b_i \; \sigma(\alpha$,p$_i$) where the branchings $b_i$ describe the probability that the $i^{th}$ excited state in $^{25}$Al finally decays to the ground state; $(1 - b_i)$ corresponds to the probability that the $i^{th}$ state decays by proton emission. In practice, $b_i = 1$ for all states below the proton binding energy in $^{25}$Al of 2.27 MeV, and $b_i \approx 0$ for most states above 2.27~MeV; only a few states above 2.27~MeV with high $J^\pi$ decay preferentially by $\gamma$-emission to the ground state of $^{25}$Al. It has been seen from the present study that such manual summing of the different exit channels in combination with the dominating ($\alpha,p$) and ($\alpha,2p$) channels makes the calculation almost insensitive to other ingredients of the statistical model like the gamma-ray strength function, the level density, or the proton optical model potential. This theoretical cross section reproduces the energy dependence of the new experimental data quite well. For the best reproduction of the new data, the theoretical $^{25}$Al production cross sections have been scaled down by a factor of 0.93 which is well within the expected uncertainty of a factor of two. As seen in Fig. \ref{fig:CS}, for the lowest center-of-mass energies ($\lesssim$2.5~MeV), these TALYS cross sections are similar to those from NON-SMOKER divided by a factor of two, but differ at higher energies due to a larger contribution of the ($\alpha,2p$) channel which is not subtracted in NON-SMOKER.

Some observed deviations between theory and experiment could be explained by the presence of strong resonances or due to a higher level density. For instance, the two cross section data points at 5.19 and 5.55 MeV covering center-of-mass energies of 4.98--5.74~MeV show deviations from the TALYS predictions. A rough correspondence to resonances in $^{22}$Ne($\alpha,n$)$^{25}$Mg mirror reaction \cite{Haas1973,Drotleff1993} can be found, taking into account an energy shift of $\sim$300 keV between mirror states in the $^{26}$Mg and $^{26}$Si as determined by Matic \textit{et al.} \cite{Matic2010} and Seweryniak \textit{et al.} \cite{Seweryniak2007}.

\begin{figure}[ht]
\includegraphics[width=0.5\textwidth]{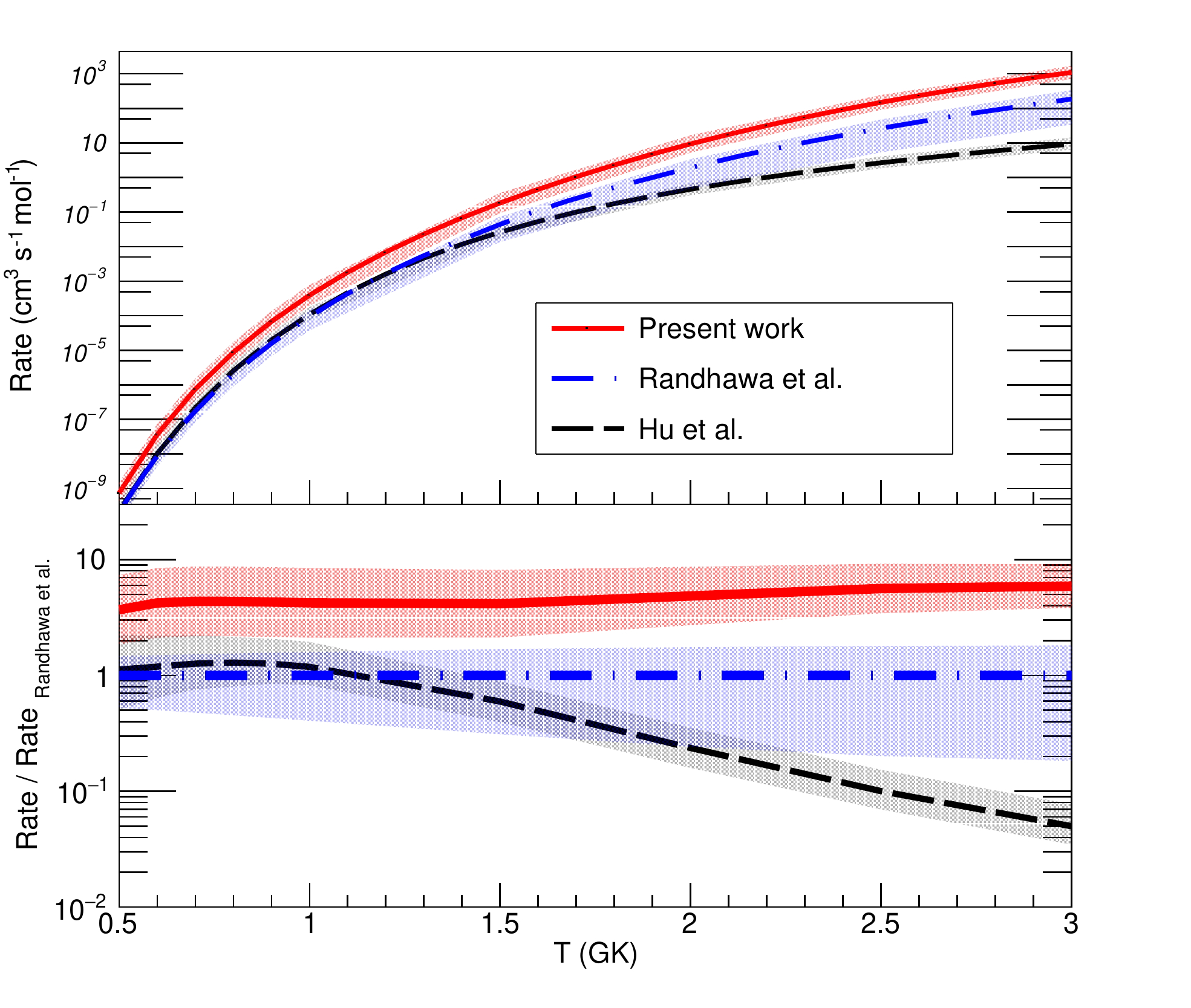}
\caption{\label{fig:rate} (upper panel) The \reac reaction rate based on this work in comparison to the rates from Randhawa \textit{et al.} \cite{Randhawa2020} and Hu \textit{et al.} \cite{Hu2021}. (lower panel) The same reaction rates as a ratio to the rate by Randhawa \textit{et al.} \cite{Randhawa2020}.}
\end{figure}

The present work provides an experimental reaction rate corresponding to the upper end of the Gamow window for T$\approx$1.8~GK. To extrapolate the rate to lower temperatures, the total reaction cross sections from the present measurement were combined with the TALYS predictions at lower energies with an uncertainty of a factor of 2. The resulting astrophysical reaction rate was calculated using the code Exp2Rate \cite{Rauscher}, and is shown in Fig. \ref{fig:rate} (red solid line) along with the rates from Randhawa \textit{et al.} \cite{Randhawa2020} (blue dot dashed line) and Hu \textit{et al.} \cite{Hu2021} (black dashed line). 
The best fitted parameters used to obtain the reaction rate in the seven-parameter REACLIB format \cite{REACLIB} in the temperature range of 0.4-2~GK can be given by $a_0$ = 46.225, $a_1$ = 0, $a_2$ = -53.08, $a_3$ = -1.035, $a_4$ = -0.227, $a_5$ = 0.116 and $a_6$ = -0.67.
As can be seen in Fig. \ref{fig:rate}, the \reac reaction rate from the present work is significantly higher than those by Refs. \cite{Randhawa2020} and \cite{Hu2021}. The factor of $\approx$4 between the new \reac rate and that of Randhawa \ea reflects the factor of $\approx$4 between the reaction cross sections for E$_{cm}\leq$ 3~MeV. Although not understood, this discrepancy could be associated to the model dependency of the obtained angle-integrated cross section in Ref. \cite{Randhawa2020}.
The discrepancy with the measurement of Ref. \cite{Hu2021} could be explained by the low number of states measured in their work. As discussed before, when compared to the mirror reaction, the number of candidate levels which could potentially contribute is expected to be about a factor of $\sim$10 higher. A possible reason could be that there might be resonances in \Si which preferentially decay by proton emission to excited states in $^{25}$Al that would only show a weak signal in the excitation curve of proton elastic scattering. Hence, resonances with small $\Gamma_{p0}$ in \Si may remain below the detection limit of this measurement. Additionally, the 2-MeV excitation energy interval measured by Ref. \cite{Hu2021} in \Si only covers the Gamow window up to $\sim$1 GK, which explains the larger discrepancy at higher temperatures.

The implications of the new \reac reaction rate for XRB model calculations and the flow into the $\alpha$p-process, is discussed below. Since the \reac reaction rate from the present work has amplitudes between that calculated using the NON-SMOKER cross sections and those of Refs. \cite{Randhawa2020} and \cite{Hu2021}, the impact of the rate from the present work on multi-zone calculations, and therefore on inferred astrophysical properties such as the light curve will be similar to those shown by Refs. \cite{Randhawa2020} and \cite{Hu2021}. But the comparatively higher rate from the present work significantly affects the nucleosynthesis reaction flow through the $^{22}$Mg waiting point in XRBs providing more insight into the underlying physics. The impact on the flow into the $\alpha$p-process can be calculated by using the two reactions affecting the destruction of $^{22}$Mg; namely \reac and $^{23}$Al($p,\gamma$)$^{24}$Si as previously described. The flow can then be defined as $\lambda_{(\alpha,p)}/(\lambda_{(\alpha,p)} + \lambda_{(p,\gamma)})$, where $\lambda_i = W_i \rho N_A\langle\sigma\nu\rangle X_{fuel}/A_{fuel}$ with $A_{fuel}$ and $X_{fuel}$ being the mass number and the mass fraction of hydrogen and helium for the $^{23}$Al($p,\gamma$)$^{24}$Si and $^{22}$Mg($\alpha,p$)$^{25}$Al reactions, respectively \cite{Merz2021}. Here, $W_i$ is a weight factor which determines the equilibrium abundance of a nuclide calculated using the Saha equation for a given temperature. In order to obtain the relevant temperatures and mass fractions, the model by Merz \& Meisel \cite{Merz2021} was adopted, where the observed features of the x-ray clockburster \textit{GS 1826-24} \cite{Ubertini1999} were reproduced for an ignition occurring at T = 0.7~GK with X$_{\rm{H}}$ = 0.06 and X$_{\rm{He}}$ = 0.19 and a peak temperature ($T_{peak}$) at 1.0~GK. The nucleosynthesis flow into the $\alpha$p-process for temperatures ranging from the ignition point to peak temperatures are of interest.

\begin{figure}[!ht]
\includegraphics[width=0.5\textwidth]{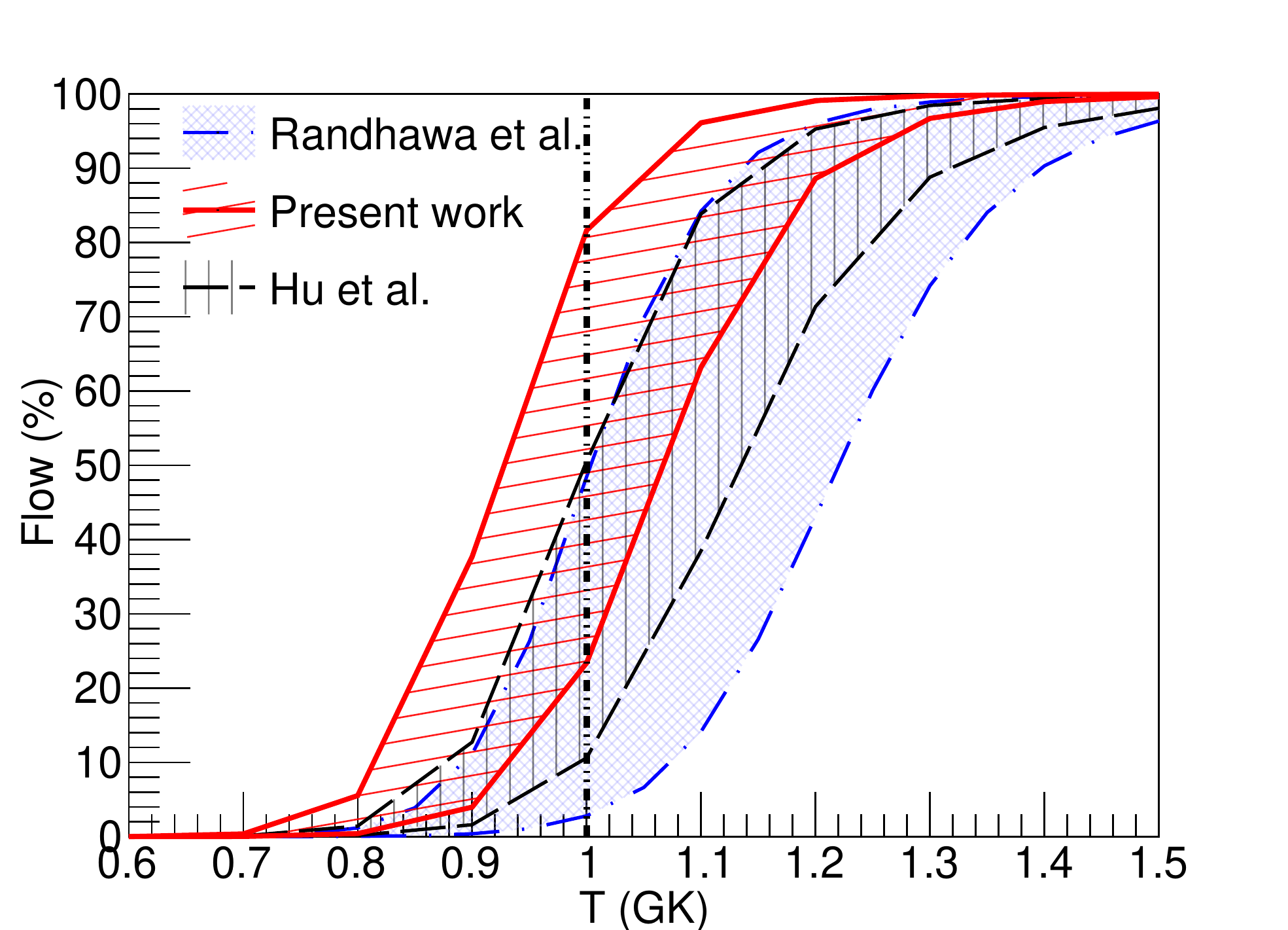}
\caption{\label{fig:Flow} The upper and lower bounds of the flow into the $\alpha$p-process calculated using the ignition conditions of Merz \& Meisel \cite{Merz2021} (X$_{\rm{H}}$ = 0.06, X$_{\rm{He}}$ = 0.19) using the \reac rates from the present work (red solid lines), Hu \ea (black dashed line) and Randhawa \ea (blue dot-dashed line). The vertical line denotes the peak temperature for the model by Ref. \cite{Merz2021}.}
\end{figure}

Fig. \ref{fig:Flow} shows the nucleosynthesis flow into the $\alpha$p-process calculated using the \reac reaction rates from the present work, Randhawa \ea \cite{Randhawa2020} and Hu \ea \cite{Hu2021} for the ignition conditions by Ref. \cite{Merz2021}. The $^{23}$Al($p,\gamma$)$^{24}$Si reaction rate for these calculations are from a recent high-precision mass measurement of $^{24}$Si which significantly reduced the previous rate uncertainties \cite{Puentes2022}. For the following discussion, this work will consider a 10\% flow as the onset point or the point when the $\alpha$p-process becomes significant (as was done in Ref. \cite{Puentes2022}) and will consider a flow greater than 50\% as a significant bypass of the $^{22}$Mg waiting point via the $(\alpha,p)$ reaction. As seen in Fig. \ref{fig:Flow}, for the temperatures found to be relevant for the XRB calculations by Merz \& Meisel (T = 0.7-1.0~GK), the previous \reac rates by Randhawa \ea and Hu \ea results in a relatively minor flow into the $\alpha$p-process at $^{22}$Mg (ranging from 3\% to 51\% even at $T_{peak}$ = 1~GK), indicating that the $^{22}$Mg waiting point is potentially not bypassed. Using the results from the present work, the flow into the $\alpha$p-process is significant with at least 23\% and as high as 82\% at 1~GK. In addition, this new rate suggests that the onset of the $\alpha$p-process occurs at a lower temperature $T_{9}=0.87\pm{0.06}$, while the onset temperatures using the rates of Randhawa \ea and Hu \ea occurs at  $T_{9}=0.99\pm{0.09}$ and $T_{9}=0.94\pm{0.06}$, respectively, which is close to the peak temperature.

This work presents a new direct measurement of the angle- and energy- integrated cross sections of the \reac reaction, which was found to be a factor of $\approx$ 4 higher than the previous direct measurement. The new reaction rate presented in this work shows a significant nucleosynthesis flow into the $\alpha$p-process at the $^{22}$Mg waiting point, contradicting recent results which found this to be relatively minor with the main nucleosynthesis flow occuring via $^{22}$Mg($p,\gamma$)$^{23}$Al($p,\gamma$)$^{24}$Si. In addition, this work found that the onset of the \reac reaction occurs at lower temperatures.


This material is based upon work supported by the U.S. Department of Energy, Office of Science, Office of Nuclear Physics, under contract number DE-AC02-06CH11357, and National Research Development and Innovation Office (NKFIH), Budapest, Hungary (K134197). This research used resources of Argonne National Laboratory's ATLAS facility, which is a DOE Office of Science User Facility. AP also acknowledges support from the Deutsche Forschungsgemeinschaft (DFG, German Research Foundation)-Project No. 279384907-SFB 1245, and the State of Hesse within the Research Cluster ELEMENTS (Project ID 500/10.006).

\normalem
\bibliography{apssamp}

\begin{thebibliography}{40}%
\makeatletter
\providecommand \@ifxundefined [1]{%
 \@ifx{#1\undefined}
}%
\providecommand \@ifnum [1]{%
 \ifnum #1\expandafter \@firstoftwo
 \else \expandafter \@secondoftwo
 \fi
}%
\providecommand \@ifx [1]{%
 \ifx #1\expandafter \@firstoftwo
 \else \expandafter \@secondoftwo
 \fi
}%
\providecommand \natexlab [1]{#1}%
\providecommand \enquote  [1]{``#1''}%
\providecommand \bibnamefont  [1]{#1}%
\providecommand \bibfnamefont [1]{#1}%
\providecommand \citenamefont [1]{#1}%
\providecommand \href@noop [0]{\@secondoftwo}%
\providecommand \href [0]{\begingroup \@sanitize@url \@href}%
\providecommand \@href[1]{\@@startlink{#1}\@@href}%
\providecommand \@@href[1]{\endgroup#1\@@endlink}%
\providecommand \@sanitize@url [0]{\catcode `\\12\catcode `\$12\catcode
  `\&12\catcode `\#12\catcode `\^12\catcode `\_12\catcode `\%12\relax}%
\providecommand \@@startlink[1]{}%
\providecommand \@@endlink[0]{}%
\providecommand \url  [0]{\begingroup\@sanitize@url \@url }%
\providecommand \@url [1]{\endgroup\@href {#1}{\urlprefix }}%
\providecommand \urlprefix  [0]{URL }%
\providecommand \Eprint [0]{\href }%
\providecommand \doibase [0]{https://doi.org/}%
\providecommand \selectlanguage [0]{\@gobble}%
\providecommand \bibinfo  [0]{\@secondoftwo}%
\providecommand \bibfield  [0]{\@secondoftwo}%
\providecommand \translation [1]{[#1]}%
\providecommand \BibitemOpen [0]{}%
\providecommand \bibitemStop [0]{}%
\providecommand \bibitemNoStop [0]{.\EOS\space}%
\providecommand \EOS [0]{\spacefactor3000\relax}%
\providecommand \BibitemShut  [1]{\csname bibitem#1\endcsname}%
\let\auto@bib@innerbib\@empty
\bibitem [{\citenamefont {{Lewin}}\ \emph {et~al.}(1993)\citenamefont
  {{Lewin}}, \citenamefont {{van Paradijs}},\ and\ \citenamefont
  {{Taam}}}]{Lewin1993}%
  \BibitemOpen
  \bibfield  {author} {\bibinfo {author} {\bibfnamefont {W.~H.~G.}\
  \bibnamefont {{Lewin}}}, \bibinfo {author} {\bibfnamefont {J.}~\bibnamefont
  {{van Paradijs}}},\ and\ \bibinfo {author} {\bibfnamefont {R.~E.}\
  \bibnamefont {{Taam}}},\ }\href {https://doi.org/10.1007/BF00196124}
  {\bibfield  {journal} {\bibinfo  {journal} {Space Sci. Rev.}\ }\textbf
  {\bibinfo {volume} {62}},\ \bibinfo {pages} {223} (\bibinfo {year}
  {1993})}\BibitemShut {NoStop}%
\bibitem [{\citenamefont {{Schatz}}\ and\ \citenamefont
  {{Rehm}}(2006)}]{Schatz2006}%
  \BibitemOpen
  \bibfield  {author} {\bibinfo {author} {\bibfnamefont {H.}~\bibnamefont
  {{Schatz}}}\ and\ \bibinfo {author} {\bibfnamefont {K.~E.}\ \bibnamefont
  {{Rehm}}},\ }\href {https://doi.org/10.1016/j.nuclphysa.2005.05.200}
  {\bibfield  {journal} {\bibinfo  {journal} {Nucl. Phys. A}\ }\textbf
  {\bibinfo {volume} {777}},\ \bibinfo {pages} {601} (\bibinfo {year}
  {2006})}\BibitemShut {NoStop}%
\bibitem [{\citenamefont {{Parikh}}\ \emph {et~al.}(2013)\citenamefont
  {{Parikh}}, \citenamefont {{Jos{\'e}}}, \citenamefont {{Sala}},\ and\
  \citenamefont {{Iliadis}}}]{Parikh2013}%
  \BibitemOpen
  \bibfield  {author} {\bibinfo {author} {\bibfnamefont {A.}~\bibnamefont
  {{Parikh}}}, \bibinfo {author} {\bibfnamefont {J.}~\bibnamefont
  {{Jos{\'e}}}}, \bibinfo {author} {\bibfnamefont {G.}~\bibnamefont {{Sala}}},\
  and\ \bibinfo {author} {\bibfnamefont {C.}~\bibnamefont {{Iliadis}}},\ }\href
  {https://doi.org/10.1016/j.ppnp.2012.11.002} {\bibfield  {journal} {\bibinfo
  {journal} {Prog. Part. Nucl. Phys.}\ }\textbf {\bibinfo {volume} {69}},\
  \bibinfo {pages} {225} (\bibinfo {year} {2013})}\BibitemShut {NoStop}%
\bibitem [{\citenamefont {Schatz}\ \emph {et~al.}(1998)\citenamefont {Schatz},
  \citenamefont {Aprahamian}, \citenamefont {Görres}, \citenamefont
  {Wiescher}, \citenamefont {Rauscher}, \citenamefont {Rembges}, \citenamefont
  {Thielemann}, \citenamefont {Pfeiffer}, \citenamefont {Möller},
  \citenamefont {Kratz}, \citenamefont {Herndl}, \citenamefont {Brown},\ and\
  \citenamefont {Rebel}}]{Schatz1998}%
  \BibitemOpen
  \bibfield  {author} {\bibinfo {author} {\bibfnamefont {H.}~\bibnamefont
  {Schatz}}, \bibinfo {author} {\bibfnamefont {A.}~\bibnamefont {Aprahamian}},
  \bibinfo {author} {\bibfnamefont {J.}~\bibnamefont {Görres}}, \bibinfo
  {author} {\bibfnamefont {M.}~\bibnamefont {Wiescher}}, \bibinfo {author}
  {\bibfnamefont {T.}~\bibnamefont {Rauscher}}, \bibinfo {author}
  {\bibfnamefont {J.}~\bibnamefont {Rembges}}, \bibinfo {author} {\bibfnamefont
  {F.-K.}\ \bibnamefont {Thielemann}}, \bibinfo {author} {\bibfnamefont
  {B.}~\bibnamefont {Pfeiffer}}, \bibinfo {author} {\bibfnamefont
  {P.}~\bibnamefont {Möller}}, \bibinfo {author} {\bibfnamefont {K.-L.}\
  \bibnamefont {Kratz}}, \bibinfo {author} {\bibfnamefont {H.}~\bibnamefont
  {Herndl}}, \bibinfo {author} {\bibfnamefont {B.}~\bibnamefont {Brown}},\ and\
  \bibinfo {author} {\bibfnamefont {H.}~\bibnamefont {Rebel}},\ }\href
  {https://doi.org/10.1016/S0370-1573(97)00048-3} {\bibfield  {journal}
  {\bibinfo  {journal} {Phys. Rep.}\ }\textbf {\bibinfo {volume} {294}},\
  \bibinfo {pages} {167} (\bibinfo {year} {1998})}\BibitemShut {NoStop}%
\bibitem [{\citenamefont {{Fisker}}\ \emph {et~al.}(2008)\citenamefont
  {{Fisker}}, \citenamefont {{Schatz}},\ and\ \citenamefont
  {{Thielemann}}}]{Fisker2008}%
  \BibitemOpen
  \bibfield  {author} {\bibinfo {author} {\bibfnamefont {J.~L.}\ \bibnamefont
  {{Fisker}}}, \bibinfo {author} {\bibfnamefont {H.}~\bibnamefont {{Schatz}}},\
  and\ \bibinfo {author} {\bibfnamefont {F.-K.}\ \bibnamefont {{Thielemann}}},\
  }\href {https://doi.org/10.1086/521104} {\bibfield  {journal} {\bibinfo
  {journal} {Astrophys. J. Suppl. Series}\ }\textbf {\bibinfo {volume} {174}},\
  \bibinfo {pages} {261} (\bibinfo {year} {2008})}\BibitemShut {NoStop}%
\bibitem [{\citenamefont {{Meisel}}\ \emph {et~al.}(2018)\citenamefont
  {{Meisel}}, \citenamefont {{Deibel}}, \citenamefont {{Keek}}, \citenamefont
  {{Shternin}},\ and\ \citenamefont {{Elfritz}}}]{Meisel2018}%
  \BibitemOpen
  \bibfield  {author} {\bibinfo {author} {\bibfnamefont {Z.}~\bibnamefont
  {{Meisel}}}, \bibinfo {author} {\bibfnamefont {A.}~\bibnamefont {{Deibel}}},
  \bibinfo {author} {\bibfnamefont {L.}~\bibnamefont {{Keek}}}, \bibinfo
  {author} {\bibfnamefont {P.}~\bibnamefont {{Shternin}}},\ and\ \bibinfo
  {author} {\bibfnamefont {J.}~\bibnamefont {{Elfritz}}},\ }\href
  {https://doi.org/10.1088/1361-6471/aad171} {\bibfield  {journal} {\bibinfo
  {journal} {J. Phys. G: Nucl. Part. Phys.}\ }\textbf {\bibinfo {volume}
  {45}},\ \bibinfo {pages} {093001} (\bibinfo {year} {2018})}\BibitemShut
  {NoStop}%
\bibitem [{\citenamefont {{Meisel}}(2018)}]{Meisel20182}%
  \BibitemOpen
  \bibfield  {author} {\bibinfo {author} {\bibfnamefont {Z.}~\bibnamefont
  {{Meisel}}},\ }\href {https://doi.org/10.3847/1538-4357/aac3d3} {\bibfield
  {journal} {\bibinfo  {journal} {Astrophys. J.}\ }\textbf {\bibinfo {volume}
  {860}},\ \bibinfo {eid} {147} (\bibinfo {year} {2018})}\BibitemShut {NoStop}%
\bibitem [{\citenamefont {Cyburt}\ \emph {et~al.}(2016)\citenamefont {Cyburt},
  \citenamefont {Amthor}, \citenamefont {Heger}, \citenamefont {Johnson},
  \citenamefont {Keek}, \citenamefont {Meisel}, \citenamefont {Schatz},\ and\
  \citenamefont {Smith}}]{Cyburt2016}%
  \BibitemOpen
  \bibfield  {author} {\bibinfo {author} {\bibfnamefont {R.~H.}\ \bibnamefont
  {Cyburt}}, \bibinfo {author} {\bibfnamefont {A.~M.}\ \bibnamefont {Amthor}},
  \bibinfo {author} {\bibfnamefont {A.}~\bibnamefont {Heger}}, \bibinfo
  {author} {\bibfnamefont {E.}~\bibnamefont {Johnson}}, \bibinfo {author}
  {\bibfnamefont {L.}~\bibnamefont {Keek}}, \bibinfo {author} {\bibfnamefont
  {Z.}~\bibnamefont {Meisel}}, \bibinfo {author} {\bibfnamefont
  {H.}~\bibnamefont {Schatz}},\ and\ \bibinfo {author} {\bibfnamefont
  {K.}~\bibnamefont {Smith}},\ }\href
  {https://doi.org/10.3847/0004-637X/830/2/55} {\bibfield  {journal} {\bibinfo
  {journal} {Astrophys. J.}\ }\textbf {\bibinfo {volume} {830}},\ \bibinfo
  {eid} {55} (\bibinfo {year} {2016})}\BibitemShut {NoStop}%
\bibitem [{\citenamefont {{Meisel}}\ \emph {et~al.}(2019)\citenamefont
  {{Meisel}}, \citenamefont {{Merz}},\ and\ \citenamefont
  {{Medvid}}}]{Meisel2019}%
  \BibitemOpen
  \bibfield  {author} {\bibinfo {author} {\bibfnamefont {Z.}~\bibnamefont
  {{Meisel}}}, \bibinfo {author} {\bibfnamefont {G.}~\bibnamefont {{Merz}}},\
  and\ \bibinfo {author} {\bibfnamefont {S.}~\bibnamefont {{Medvid}}},\ }\href
  {https://doi.org/10.3847/1538-4357/aafede} {\bibfield  {journal} {\bibinfo
  {journal} {Astrophys. J.}\ }\textbf {\bibinfo {volume} {872}},\ \bibinfo
  {eid} {84} (\bibinfo {year} {2019})}\BibitemShut {NoStop}%
\bibitem [{\citenamefont {Wiescher}\ \emph {et~al.}(1999)\citenamefont
  {Wiescher}, \citenamefont {Görres},\ and\ \citenamefont
  {Schatz}}]{Wiescher1999}%
  \BibitemOpen
  \bibfield  {author} {\bibinfo {author} {\bibfnamefont {M.}~\bibnamefont
  {Wiescher}}, \bibinfo {author} {\bibfnamefont {J.}~\bibnamefont {Görres}},\
  and\ \bibinfo {author} {\bibfnamefont {H.}~\bibnamefont {Schatz}},\ }\href
  {https://doi.org/10.1088/0954-3899/25/6/201} {\bibfield  {journal} {\bibinfo
  {journal} {J. Phys. G: Nucl. Part. Phys.}\ }\textbf {\bibinfo {volume}
  {25}},\ \bibinfo {pages} {R133} (\bibinfo {year} {1999})}\BibitemShut
  {NoStop}%
\bibitem [{\citenamefont {{Fisker}}\ \emph {et~al.}(2004)\citenamefont
  {{Fisker}}, \citenamefont {{Thielemann}},\ and\ \citenamefont
  {{Wiescher}}}]{Fisker2004}%
  \BibitemOpen
  \bibfield  {author} {\bibinfo {author} {\bibfnamefont {J.~L.}\ \bibnamefont
  {{Fisker}}}, \bibinfo {author} {\bibfnamefont {F.-K.}\ \bibnamefont
  {{Thielemann}}},\ and\ \bibinfo {author} {\bibfnamefont {M.}~\bibnamefont
  {{Wiescher}}},\ }\href {https://doi.org/10.1086/422215} {\bibfield  {journal}
  {\bibinfo  {journal} {Astrophys. J. Lett.}\ }\textbf {\bibinfo {volume}
  {608}},\ \bibinfo {pages} {L61} (\bibinfo {year} {2004})}\BibitemShut
  {NoStop}%
\bibitem [{\citenamefont {Basunia}(2015)}]{BASUNIA2015}%
  \BibitemOpen
  \bibfield  {author} {\bibinfo {author} {\bibfnamefont {M.~S.}\ \bibnamefont
  {Basunia}},\ }\href
  {https://doi.org/https://doi.org/10.1016/j.nds.2015.07.002} {\bibfield
  {journal} {\bibinfo  {journal} {Nuclear Data Sheets}\ }\textbf {\bibinfo
  {volume} {127}},\ \bibinfo {pages} {69} (\bibinfo {year} {2015})}\BibitemShut
  {NoStop}%
\bibitem [{\citenamefont {{Huang}}\ \emph {et~al.}(2021)\citenamefont
  {{Huang}}, \citenamefont {{Wang}}, \citenamefont {{Kondev}}, \citenamefont
  {{Audi}},\ and\ \citenamefont {{Naimi}}}]{Huang2021}%
  \BibitemOpen
  \bibfield  {author} {\bibinfo {author} {\bibfnamefont {W.~J.}\ \bibnamefont
  {{Huang}}}, \bibinfo {author} {\bibfnamefont {M.}~\bibnamefont {{Wang}}},
  \bibinfo {author} {\bibfnamefont {F.~G.}\ \bibnamefont {{Kondev}}}, \bibinfo
  {author} {\bibfnamefont {G.}~\bibnamefont {{Audi}}},\ and\ \bibinfo {author}
  {\bibfnamefont {S.}~\bibnamefont {{Naimi}}},\ }\href
  {https://doi.org/10.1088/1674-1137/abddb0} {\bibfield  {journal} {\bibinfo
  {journal} {Chinese Physics C}\ }\textbf {\bibinfo {volume} {45}},\ \bibinfo
  {eid} {030002} (\bibinfo {year} {2021})}\BibitemShut {NoStop}%
\bibitem [{\citenamefont {Puentes}\ \emph {et~al.}(2022)\citenamefont
  {Puentes}, \citenamefont {Meisel}, \citenamefont {Bollen}, \citenamefont
  {Hamaker}, \citenamefont {Langer}, \citenamefont {Leistenschneider},
  \citenamefont {Nicoloff}, \citenamefont {Ong}, \citenamefont {Redshaw},
  \citenamefont {Ringle}, \citenamefont {Sumithrarachchi}, \citenamefont
  {Surbrook}, \citenamefont {Valverde},\ and\ \citenamefont
  {Yandow}}]{Puentes2022}%
  \BibitemOpen
  \bibfield  {author} {\bibinfo {author} {\bibfnamefont {D.}~\bibnamefont
  {Puentes}}, \bibinfo {author} {\bibfnamefont {Z.}~\bibnamefont {Meisel}},
  \bibinfo {author} {\bibfnamefont {G.}~\bibnamefont {Bollen}}, \bibinfo
  {author} {\bibfnamefont {A.}~\bibnamefont {Hamaker}}, \bibinfo {author}
  {\bibfnamefont {C.}~\bibnamefont {Langer}}, \bibinfo {author} {\bibfnamefont
  {E.}~\bibnamefont {Leistenschneider}}, \bibinfo {author} {\bibfnamefont
  {C.}~\bibnamefont {Nicoloff}}, \bibinfo {author} {\bibfnamefont {W.-J.}\
  \bibnamefont {Ong}}, \bibinfo {author} {\bibfnamefont {M.}~\bibnamefont
  {Redshaw}}, \bibinfo {author} {\bibfnamefont {R.}~\bibnamefont {Ringle}},
  \bibinfo {author} {\bibfnamefont {C.~S.}\ \bibnamefont {Sumithrarachchi}},
  \bibinfo {author} {\bibfnamefont {J.}~\bibnamefont {Surbrook}}, \bibinfo
  {author} {\bibfnamefont {A.~A.}\ \bibnamefont {Valverde}},\ and\ \bibinfo
  {author} {\bibfnamefont {I.~T.}\ \bibnamefont {Yandow}},\ }\href
  {https://doi.org/10.1103/PhysRevC.106.L012801} {\bibfield  {journal}
  {\bibinfo  {journal} {\prc}\ }\textbf {\bibinfo {volume} {106}},\ \bibinfo
  {eid} {L012801} (\bibinfo {year} {2022})}\BibitemShut {NoStop}%
\bibitem [{\citenamefont {Lotay}\ \emph {et~al.}(2022)\citenamefont {Lotay},
  \citenamefont {Henderson}, \citenamefont {Catford}, \citenamefont {Ali},
  \citenamefont {Berean}, \citenamefont {Bernier}, \citenamefont
  {Bhattacharjee}, \citenamefont {Bowry}, \citenamefont {Caballero-Folch},
  \citenamefont {Davids}, \citenamefont {Drake}, \citenamefont {Garnsworthy},
  \citenamefont {{Ghazi Moradi}}, \citenamefont {Gillespie}, \citenamefont
  {Greaves}, \citenamefont {Hackman}, \citenamefont {Hallam}, \citenamefont
  {Hymers}, \citenamefont {Kasanda}, \citenamefont {Levy}, \citenamefont
  {Luna}, \citenamefont {Mathews}, \citenamefont {Meisel}, \citenamefont
  {Moukaddam}, \citenamefont {Muecher}, \citenamefont {Olaizola}, \citenamefont
  {Orr}, \citenamefont {Patel}, \citenamefont {Rajabali}, \citenamefont
  {Saito}, \citenamefont {Smallcombe}, \citenamefont {Spencer}, \citenamefont
  {Svensson}, \citenamefont {Whitmore},\ and\ \citenamefont
  {Williams}}]{Lotay2022}%
  \BibitemOpen
  \bibfield  {author} {\bibinfo {author} {\bibfnamefont {G.}~\bibnamefont
  {Lotay}}, \bibinfo {author} {\bibfnamefont {J.}~\bibnamefont {Henderson}},
  \bibinfo {author} {\bibfnamefont {W.}~\bibnamefont {Catford}}, \bibinfo
  {author} {\bibfnamefont {F.}~\bibnamefont {Ali}}, \bibinfo {author}
  {\bibfnamefont {J.}~\bibnamefont {Berean}}, \bibinfo {author} {\bibfnamefont
  {N.}~\bibnamefont {Bernier}}, \bibinfo {author} {\bibfnamefont
  {S.}~\bibnamefont {Bhattacharjee}}, \bibinfo {author} {\bibfnamefont
  {M.}~\bibnamefont {Bowry}}, \bibinfo {author} {\bibfnamefont
  {R.}~\bibnamefont {Caballero-Folch}}, \bibinfo {author} {\bibfnamefont
  {B.}~\bibnamefont {Davids}}, \bibinfo {author} {\bibfnamefont
  {T.}~\bibnamefont {Drake}}, \bibinfo {author} {\bibfnamefont
  {A.}~\bibnamefont {Garnsworthy}}, \bibinfo {author} {\bibfnamefont
  {F.}~\bibnamefont {{Ghazi Moradi}}}, \bibinfo {author} {\bibfnamefont
  {S.}~\bibnamefont {Gillespie}}, \bibinfo {author} {\bibfnamefont
  {B.}~\bibnamefont {Greaves}}, \bibinfo {author} {\bibfnamefont
  {G.}~\bibnamefont {Hackman}}, \bibinfo {author} {\bibfnamefont
  {S.}~\bibnamefont {Hallam}}, \bibinfo {author} {\bibfnamefont
  {D.}~\bibnamefont {Hymers}}, \bibinfo {author} {\bibfnamefont
  {E.}~\bibnamefont {Kasanda}}, \bibinfo {author} {\bibfnamefont
  {D.}~\bibnamefont {Levy}}, \bibinfo {author} {\bibfnamefont {B.}~\bibnamefont
  {Luna}}, \bibinfo {author} {\bibfnamefont {A.}~\bibnamefont {Mathews}},
  \bibinfo {author} {\bibfnamefont {Z.}~\bibnamefont {Meisel}}, \bibinfo
  {author} {\bibfnamefont {M.}~\bibnamefont {Moukaddam}}, \bibinfo {author}
  {\bibfnamefont {D.}~\bibnamefont {Muecher}}, \bibinfo {author} {\bibfnamefont
  {B.}~\bibnamefont {Olaizola}}, \bibinfo {author} {\bibfnamefont
  {N.}~\bibnamefont {Orr}}, \bibinfo {author} {\bibfnamefont {H.}~\bibnamefont
  {Patel}}, \bibinfo {author} {\bibfnamefont {M.}~\bibnamefont {Rajabali}},
  \bibinfo {author} {\bibfnamefont {Y.}~\bibnamefont {Saito}}, \bibinfo
  {author} {\bibfnamefont {J.}~\bibnamefont {Smallcombe}}, \bibinfo {author}
  {\bibfnamefont {M.}~\bibnamefont {Spencer}}, \bibinfo {author} {\bibfnamefont
  {C.}~\bibnamefont {Svensson}}, \bibinfo {author} {\bibfnamefont
  {K.}~\bibnamefont {Whitmore}},\ and\ \bibinfo {author} {\bibfnamefont
  {M.}~\bibnamefont {Williams}},\ }\href
  {https://doi.org/10.1016/j.physletb.2022.137361} {\bibfield  {journal}
  {\bibinfo  {journal} {Phys. Lett. B}\ }\textbf {\bibinfo {volume} {833}},\
  \bibinfo {eid} {137361} (\bibinfo {year} {2022})}\BibitemShut {NoStop}%
\bibitem [{\citenamefont {Cyburt}\ \emph {et~al.}(2010)\citenamefont {Cyburt},
  \citenamefont {Amthor}, \citenamefont {Ferguson}, \citenamefont {Meisel},
  \citenamefont {Smith}, \citenamefont {Warren}, \citenamefont {Heger},
  \citenamefont {Hoffman}, \citenamefont {Rauscher}, \citenamefont {Sakharuk},
  \citenamefont {Schatz}, \citenamefont {Thielemann},\ and\ \citenamefont
  {Wiescher}}]{Cyburt2010}%
  \BibitemOpen
  \bibfield  {author} {\bibinfo {author} {\bibfnamefont {R.~H.}\ \bibnamefont
  {Cyburt}}, \bibinfo {author} {\bibfnamefont {A.~M.}\ \bibnamefont {Amthor}},
  \bibinfo {author} {\bibfnamefont {R.}~\bibnamefont {Ferguson}}, \bibinfo
  {author} {\bibfnamefont {Z.}~\bibnamefont {Meisel}}, \bibinfo {author}
  {\bibfnamefont {K.}~\bibnamefont {Smith}}, \bibinfo {author} {\bibfnamefont
  {S.}~\bibnamefont {Warren}}, \bibinfo {author} {\bibfnamefont
  {A.}~\bibnamefont {Heger}}, \bibinfo {author} {\bibfnamefont {R.~D.}\
  \bibnamefont {Hoffman}}, \bibinfo {author} {\bibfnamefont {T.}~\bibnamefont
  {Rauscher}}, \bibinfo {author} {\bibfnamefont {A.}~\bibnamefont {Sakharuk}},
  \bibinfo {author} {\bibfnamefont {H.}~\bibnamefont {Schatz}}, \bibinfo
  {author} {\bibfnamefont {F.~K.}\ \bibnamefont {Thielemann}},\ and\ \bibinfo
  {author} {\bibfnamefont {M.}~\bibnamefont {Wiescher}},\ }\href
  {https://doi.org/10.1088/0067-0049/189/1/240} {\bibfield  {journal} {\bibinfo
   {journal} {Astrophys. J. Suppl. Series}\ }\textbf {\bibinfo {volume}
  {189}},\ \bibinfo {pages} {240} (\bibinfo {year} {2010})}\BibitemShut
  {NoStop}%
\bibitem [{\citenamefont {Matic}\ \emph {et~al.}(2011)\citenamefont {Matic},
  \citenamefont {van~den Berg}, \citenamefont {Harakeh}, \citenamefont
  {W\"ortche}, \citenamefont {Beard}, \citenamefont {Berg}, \citenamefont
  {G\"orres}, \citenamefont {LeBlanc}, \citenamefont {O'Brien}, \citenamefont
  {Wiescher}, \citenamefont {Fujita}, \citenamefont {Hatanaka}, \citenamefont
  {Sakemi}, \citenamefont {Shimizu}, \citenamefont {Tameshige}, \citenamefont
  {Tamii}, \citenamefont {Yosoi}, \citenamefont {Adachi}, \citenamefont
  {Fujita}, \citenamefont {Shimbara}, \citenamefont {Fujita}, \citenamefont
  {Wakasa}, \citenamefont {Greene}, \citenamefont {Crowter},\ and\
  \citenamefont {Schatz}}]{Matic2011}%
  \BibitemOpen
  \bibfield  {author} {\bibinfo {author} {\bibfnamefont {A.}~\bibnamefont
  {Matic}}, \bibinfo {author} {\bibfnamefont {A.~M.}\ \bibnamefont {van~den
  Berg}}, \bibinfo {author} {\bibfnamefont {M.~N.}\ \bibnamefont {Harakeh}},
  \bibinfo {author} {\bibfnamefont {H.~J.}\ \bibnamefont {W\"ortche}}, \bibinfo
  {author} {\bibfnamefont {M.}~\bibnamefont {Beard}}, \bibinfo {author}
  {\bibfnamefont {G.~P.~A.}\ \bibnamefont {Berg}}, \bibinfo {author}
  {\bibfnamefont {J.}~\bibnamefont {G\"orres}}, \bibinfo {author}
  {\bibfnamefont {P.}~\bibnamefont {LeBlanc}}, \bibinfo {author} {\bibfnamefont
  {S.}~\bibnamefont {O'Brien}}, \bibinfo {author} {\bibfnamefont
  {M.}~\bibnamefont {Wiescher}}, \bibinfo {author} {\bibfnamefont
  {K.}~\bibnamefont {Fujita}}, \bibinfo {author} {\bibfnamefont
  {K.}~\bibnamefont {Hatanaka}}, \bibinfo {author} {\bibfnamefont
  {Y.}~\bibnamefont {Sakemi}}, \bibinfo {author} {\bibfnamefont
  {Y.}~\bibnamefont {Shimizu}}, \bibinfo {author} {\bibfnamefont
  {Y.}~\bibnamefont {Tameshige}}, \bibinfo {author} {\bibfnamefont
  {A.}~\bibnamefont {Tamii}}, \bibinfo {author} {\bibfnamefont
  {M.}~\bibnamefont {Yosoi}}, \bibinfo {author} {\bibfnamefont
  {T.}~\bibnamefont {Adachi}}, \bibinfo {author} {\bibfnamefont
  {Y.}~\bibnamefont {Fujita}}, \bibinfo {author} {\bibfnamefont
  {Y.}~\bibnamefont {Shimbara}}, \bibinfo {author} {\bibfnamefont
  {H.}~\bibnamefont {Fujita}}, \bibinfo {author} {\bibfnamefont
  {T.}~\bibnamefont {Wakasa}}, \bibinfo {author} {\bibfnamefont {J.~P.}\
  \bibnamefont {Greene}}, \bibinfo {author} {\bibfnamefont {R.}~\bibnamefont
  {Crowter}},\ and\ \bibinfo {author} {\bibfnamefont {H.}~\bibnamefont
  {Schatz}},\ }\href {https://doi.org/10.1103/PhysRevC.84.025801} {\bibfield
  {journal} {\bibinfo  {journal} {Phys. Rev. C}\ }\textbf {\bibinfo {volume}
  {84}},\ \bibinfo {pages} {025801} (\bibinfo {year} {2011})}\BibitemShut
  {NoStop}%
\bibitem [{\citenamefont {Randhawa}\ \emph {et~al.}(2020)\citenamefont
  {Randhawa}, \citenamefont {Ayyad}, \citenamefont {Mittig}, \citenamefont
  {Meisel}, \citenamefont {Ahn}, \citenamefont {Aguilar}, \citenamefont
  {Alvarez-Pol}, \citenamefont {Bardayan}, \citenamefont {Bazin}, \citenamefont
  {Beceiro-Novo}, \citenamefont {Blankstein}, \citenamefont {Carpenter},
  \citenamefont {Cortesi}, \citenamefont {Cortina-Gil}, \citenamefont {Gastis},
  \citenamefont {Hall}, \citenamefont {Henderson}, \citenamefont {Kolata},
  \citenamefont {Mijatovic}, \citenamefont {Ndayisabye}, \citenamefont
  {O'Malley}, \citenamefont {Pereira}, \citenamefont {Pierre}, \citenamefont
  {Robert}, \citenamefont {Santamaria}, \citenamefont {Schatz}, \citenamefont
  {Smith}, \citenamefont {Watwood},\ and\ \citenamefont
  {Zamora}}]{Randhawa2020}%
  \BibitemOpen
  \bibfield  {author} {\bibinfo {author} {\bibfnamefont {J.~S.}\ \bibnamefont
  {Randhawa}}, \bibinfo {author} {\bibfnamefont {Y.}~\bibnamefont {Ayyad}},
  \bibinfo {author} {\bibfnamefont {W.}~\bibnamefont {Mittig}}, \bibinfo
  {author} {\bibfnamefont {Z.}~\bibnamefont {Meisel}}, \bibinfo {author}
  {\bibfnamefont {T.}~\bibnamefont {Ahn}}, \bibinfo {author} {\bibfnamefont
  {S.}~\bibnamefont {Aguilar}}, \bibinfo {author} {\bibfnamefont
  {H.}~\bibnamefont {Alvarez-Pol}}, \bibinfo {author} {\bibfnamefont {D.~W.}\
  \bibnamefont {Bardayan}}, \bibinfo {author} {\bibfnamefont {D.}~\bibnamefont
  {Bazin}}, \bibinfo {author} {\bibfnamefont {S.}~\bibnamefont {Beceiro-Novo}},
  \bibinfo {author} {\bibfnamefont {D.}~\bibnamefont {Blankstein}}, \bibinfo
  {author} {\bibfnamefont {L.}~\bibnamefont {Carpenter}}, \bibinfo {author}
  {\bibfnamefont {M.}~\bibnamefont {Cortesi}}, \bibinfo {author} {\bibfnamefont
  {D.}~\bibnamefont {Cortina-Gil}}, \bibinfo {author} {\bibfnamefont
  {P.}~\bibnamefont {Gastis}}, \bibinfo {author} {\bibfnamefont
  {M.}~\bibnamefont {Hall}}, \bibinfo {author} {\bibfnamefont {S.}~\bibnamefont
  {Henderson}}, \bibinfo {author} {\bibfnamefont {J.~J.}\ \bibnamefont
  {Kolata}}, \bibinfo {author} {\bibfnamefont {T.}~\bibnamefont {Mijatovic}},
  \bibinfo {author} {\bibfnamefont {F.}~\bibnamefont {Ndayisabye}}, \bibinfo
  {author} {\bibfnamefont {P.}~\bibnamefont {O'Malley}}, \bibinfo {author}
  {\bibfnamefont {J.}~\bibnamefont {Pereira}}, \bibinfo {author} {\bibfnamefont
  {A.}~\bibnamefont {Pierre}}, \bibinfo {author} {\bibfnamefont
  {H.}~\bibnamefont {Robert}}, \bibinfo {author} {\bibfnamefont
  {C.}~\bibnamefont {Santamaria}}, \bibinfo {author} {\bibfnamefont
  {H.}~\bibnamefont {Schatz}}, \bibinfo {author} {\bibfnamefont
  {J.}~\bibnamefont {Smith}}, \bibinfo {author} {\bibfnamefont
  {N.}~\bibnamefont {Watwood}},\ and\ \bibinfo {author} {\bibfnamefont {J.~C.}\
  \bibnamefont {Zamora}},\ }\href
  {https://doi.org/10.1103/PhysRevLett.125.202701} {\bibfield  {journal}
  {\bibinfo  {journal} {\prl}\ }\textbf {\bibinfo {volume} {125}},\ \bibinfo
  {eid} {202701} (\bibinfo {year} {2020})}\BibitemShut {NoStop}%
\bibitem [{\citenamefont {Koning}\ and\ \citenamefont {Rochman}(2012)}]{TALYS}%
  \BibitemOpen
  \bibfield  {author} {\bibinfo {author} {\bibfnamefont {A.}~\bibnamefont
  {Koning}}\ and\ \bibinfo {author} {\bibfnamefont {D.}~\bibnamefont
  {Rochman}},\ }\href
  {https://doi.org/https://doi.org/10.1016/j.nds.2012.11.002} {\bibfield
  {journal} {\bibinfo  {journal} {Nuclear Data Sheets}\ }\textbf {\bibinfo
  {volume} {113}},\ \bibinfo {pages} {2841} (\bibinfo {year}
  {2012})}\BibitemShut {NoStop}%
\bibitem [{NON()}]{NON_SMOKER}%
  \BibitemOpen
  \href@noop {} {}\bibinfo {note} {T Rauscher, (NON-SMOKER)
  \lowercase{h}ttp://nucastro.org/ nonsmoker.html}\BibitemShut {NoStop}%
\bibitem [{\citenamefont {Hu}\ \emph {et~al.}(2021)\citenamefont {Hu},
  \citenamefont {Yamaguchi}, \citenamefont {Lam}, \citenamefont {Heger},
  \citenamefont {Kahl}, \citenamefont {Jacobs}, \citenamefont {Johnston},
  \citenamefont {Xu}, \citenamefont {Zhang}, \citenamefont {Ma}, \citenamefont
  {Ru}, \citenamefont {Liu}, \citenamefont {Liu}, \citenamefont {Hayakawa},
  \citenamefont {Yang}, \citenamefont {Shimizu}, \citenamefont {Hamill},
  \citenamefont {Murphy}, \citenamefont {Su}, \citenamefont {Fang},
  \citenamefont {Chae}, \citenamefont {Kwag}, \citenamefont {Cha},
  \citenamefont {Duy}, \citenamefont {Uyen}, \citenamefont {Kim}, \citenamefont
  {Pizzone}, \citenamefont {La~Cognata}, \citenamefont {Cherubini},
  \citenamefont {Romano}, \citenamefont {Tumino}, \citenamefont {Liang},
  \citenamefont {Psaltis}, \citenamefont {Sferrazza}, \citenamefont {Kim},
  \citenamefont {Li},\ and\ \citenamefont {Kubono}}]{Hu2021}%
  \BibitemOpen
  \bibfield  {author} {\bibinfo {author} {\bibfnamefont {J.}~\bibnamefont
  {Hu}}, \bibinfo {author} {\bibfnamefont {H.}~\bibnamefont {Yamaguchi}},
  \bibinfo {author} {\bibfnamefont {Y.~H.}\ \bibnamefont {Lam}}, \bibinfo
  {author} {\bibfnamefont {A.}~\bibnamefont {Heger}}, \bibinfo {author}
  {\bibfnamefont {D.}~\bibnamefont {Kahl}}, \bibinfo {author} {\bibfnamefont
  {A.~M.}\ \bibnamefont {Jacobs}}, \bibinfo {author} {\bibfnamefont
  {Z.}~\bibnamefont {Johnston}}, \bibinfo {author} {\bibfnamefont {S.~W.}\
  \bibnamefont {Xu}}, \bibinfo {author} {\bibfnamefont {N.~T.}\ \bibnamefont
  {Zhang}}, \bibinfo {author} {\bibfnamefont {S.~B.}\ \bibnamefont {Ma}},
  \bibinfo {author} {\bibfnamefont {L.~H.}\ \bibnamefont {Ru}}, \bibinfo
  {author} {\bibfnamefont {E.~Q.}\ \bibnamefont {Liu}}, \bibinfo {author}
  {\bibfnamefont {T.}~\bibnamefont {Liu}}, \bibinfo {author} {\bibfnamefont
  {S.}~\bibnamefont {Hayakawa}}, \bibinfo {author} {\bibfnamefont
  {L.}~\bibnamefont {Yang}}, \bibinfo {author} {\bibfnamefont {H.}~\bibnamefont
  {Shimizu}}, \bibinfo {author} {\bibfnamefont {C.~B.}\ \bibnamefont {Hamill}},
  \bibinfo {author} {\bibfnamefont {A.~S.~J.}\ \bibnamefont {Murphy}}, \bibinfo
  {author} {\bibfnamefont {J.}~\bibnamefont {Su}}, \bibinfo {author}
  {\bibfnamefont {X.}~\bibnamefont {Fang}}, \bibinfo {author} {\bibfnamefont
  {K.~Y.}\ \bibnamefont {Chae}}, \bibinfo {author} {\bibfnamefont {M.~S.}\
  \bibnamefont {Kwag}}, \bibinfo {author} {\bibfnamefont {S.~M.}\ \bibnamefont
  {Cha}}, \bibinfo {author} {\bibfnamefont {N.~N.}\ \bibnamefont {Duy}},
  \bibinfo {author} {\bibfnamefont {N.~K.}\ \bibnamefont {Uyen}}, \bibinfo
  {author} {\bibfnamefont {D.~H.}\ \bibnamefont {Kim}}, \bibinfo {author}
  {\bibfnamefont {R.~G.}\ \bibnamefont {Pizzone}}, \bibinfo {author}
  {\bibfnamefont {M.}~\bibnamefont {La~Cognata}}, \bibinfo {author}
  {\bibfnamefont {S.}~\bibnamefont {Cherubini}}, \bibinfo {author}
  {\bibfnamefont {S.}~\bibnamefont {Romano}}, \bibinfo {author} {\bibfnamefont
  {A.}~\bibnamefont {Tumino}}, \bibinfo {author} {\bibfnamefont
  {J.}~\bibnamefont {Liang}}, \bibinfo {author} {\bibfnamefont
  {A.}~\bibnamefont {Psaltis}}, \bibinfo {author} {\bibfnamefont
  {M.}~\bibnamefont {Sferrazza}}, \bibinfo {author} {\bibfnamefont
  {D.}~\bibnamefont {Kim}}, \bibinfo {author} {\bibfnamefont {Y.~Y.}\
  \bibnamefont {Li}},\ and\ \bibinfo {author} {\bibfnamefont {S.}~\bibnamefont
  {Kubono}},\ }\href {https://doi.org/10.1103/PhysRevLett.127.172701}
  {\bibfield  {journal} {\bibinfo  {journal} {\prl}\ }\textbf {\bibinfo
  {volume} {127}},\ \bibinfo {eid} {172701} (\bibinfo {year}
  {2021})}\BibitemShut {NoStop}%
\bibitem [{\citenamefont {{Lam}}\ \emph {et~al.}(2022)\citenamefont {{Lam}},
  \citenamefont {{Liu}}, \citenamefont {{Heger}}, \citenamefont {{Lu}},
  \citenamefont {{Jacobs}},\ and\ \citenamefont {{Johnston}}}]{Lam2022}%
  \BibitemOpen
  \bibfield  {author} {\bibinfo {author} {\bibfnamefont {Y.~H.}\ \bibnamefont
  {{Lam}}}, \bibinfo {author} {\bibfnamefont {Z.~X.}\ \bibnamefont {{Liu}}},
  \bibinfo {author} {\bibfnamefont {A.}~\bibnamefont {{Heger}}}, \bibinfo
  {author} {\bibfnamefont {N.}~\bibnamefont {{Lu}}}, \bibinfo {author}
  {\bibfnamefont {A.~M.}\ \bibnamefont {{Jacobs}}},\ and\ \bibinfo {author}
  {\bibfnamefont {Z.}~\bibnamefont {{Johnston}}},\ }\href
  {https://doi.org/10.3847/1538-4357/ac4d8b} {\bibfield  {journal} {\bibinfo
  {journal} {Astrophys. J.}\ }\textbf {\bibinfo {volume} {929}},\ \bibinfo
  {eid} {72} (\bibinfo {year} {2022})}\BibitemShut {NoStop}%
\bibitem [{\citenamefont {Hoffman}\ \emph {et~al.}(2022)\citenamefont
  {Hoffman}, \citenamefont {Tang}, \citenamefont {Avila}, \citenamefont
  {Ayyad}, \citenamefont {Brown}, \citenamefont {Chen}, \citenamefont {Chipps},
  \citenamefont {Jayatissa}, \citenamefont {Kay}, \citenamefont
  {Müller-Gatermann}, \citenamefont {Ong}, \citenamefont {Song},\ and\
  \citenamefont {Wilson}}]{RAISOR}%
  \BibitemOpen
  \bibfield  {author} {\bibinfo {author} {\bibfnamefont {C.}~\bibnamefont
  {Hoffman}}, \bibinfo {author} {\bibfnamefont {T.}~\bibnamefont {Tang}},
  \bibinfo {author} {\bibfnamefont {M.}~\bibnamefont {Avila}}, \bibinfo
  {author} {\bibfnamefont {Y.}~\bibnamefont {Ayyad}}, \bibinfo {author}
  {\bibfnamefont {K.}~\bibnamefont {Brown}}, \bibinfo {author} {\bibfnamefont
  {J.}~\bibnamefont {Chen}}, \bibinfo {author} {\bibfnamefont {K.}~\bibnamefont
  {Chipps}}, \bibinfo {author} {\bibfnamefont {H.}~\bibnamefont {Jayatissa}},
  \bibinfo {author} {\bibfnamefont {B.}~\bibnamefont {Kay}}, \bibinfo {author}
  {\bibfnamefont {C.}~\bibnamefont {Müller-Gatermann}}, \bibinfo {author}
  {\bibfnamefont {H.}~\bibnamefont {Ong}}, \bibinfo {author} {\bibfnamefont
  {J.}~\bibnamefont {Song}},\ and\ \bibinfo {author} {\bibfnamefont
  {G.}~\bibnamefont {Wilson}},\ }\href
  {https://doi.org/10.1016/j.nima.2022.166612} {\bibfield  {journal} {\bibinfo
  {journal} {Nucl. Instrum. Methods Phys. Res., Sect. A}\ }\textbf {\bibinfo
  {volume} {1032}},\ \bibinfo {eid} {166612} (\bibinfo {year}
  {2022})}\BibitemShut {NoStop}%
\bibitem [{\citenamefont {Carnelli}\ \emph {et~al.}(2015)\citenamefont
  {Carnelli}, \citenamefont {Almaraz-Calderon}, \citenamefont {Rehm},
  \citenamefont {Albers}, \citenamefont {Alcorta}, \citenamefont {Bertone},
  \citenamefont {Digiovine}, \citenamefont {Esbensen}, \citenamefont
  {{Fernández Niello}}, \citenamefont {Henderson}, \citenamefont {Jiang},
  \citenamefont {Lai}, \citenamefont {Marley}, \citenamefont {Nusair},
  \citenamefont {Palchan-Hazan}, \citenamefont {Pardo}, \citenamefont {Paul},\
  and\ \citenamefont {Ugalde}}]{MUSIC_NIM}%
  \BibitemOpen
  \bibfield  {author} {\bibinfo {author} {\bibfnamefont {P.}~\bibnamefont
  {Carnelli}}, \bibinfo {author} {\bibfnamefont {S.}~\bibnamefont
  {Almaraz-Calderon}}, \bibinfo {author} {\bibfnamefont {K.}~\bibnamefont
  {Rehm}}, \bibinfo {author} {\bibfnamefont {M.}~\bibnamefont {Albers}},
  \bibinfo {author} {\bibfnamefont {M.}~\bibnamefont {Alcorta}}, \bibinfo
  {author} {\bibfnamefont {P.}~\bibnamefont {Bertone}}, \bibinfo {author}
  {\bibfnamefont {B.}~\bibnamefont {Digiovine}}, \bibinfo {author}
  {\bibfnamefont {H.}~\bibnamefont {Esbensen}}, \bibinfo {author}
  {\bibfnamefont {J.}~\bibnamefont {{Fernández Niello}}}, \bibinfo {author}
  {\bibfnamefont {D.}~\bibnamefont {Henderson}}, \bibinfo {author}
  {\bibfnamefont {C.}~\bibnamefont {Jiang}}, \bibinfo {author} {\bibfnamefont
  {J.}~\bibnamefont {Lai}}, \bibinfo {author} {\bibfnamefont {S.}~\bibnamefont
  {Marley}}, \bibinfo {author} {\bibfnamefont {O.}~\bibnamefont {Nusair}},
  \bibinfo {author} {\bibfnamefont {T.}~\bibnamefont {Palchan-Hazan}}, \bibinfo
  {author} {\bibfnamefont {R.}~\bibnamefont {Pardo}}, \bibinfo {author}
  {\bibfnamefont {M.}~\bibnamefont {Paul}},\ and\ \bibinfo {author}
  {\bibfnamefont {C.}~\bibnamefont {Ugalde}},\ }\href
  {https://www.sciencedirect.com/science/article/pii/S0168900215008591}
  {\bibfield  {journal} {\bibinfo  {journal} {Nucl. Instrum. Methods Phys.
  Res., Sect. A}\ }\textbf {\bibinfo {volume} {799}},\ \bibinfo {pages} {197}
  (\bibinfo {year} {2015})}\BibitemShut {NoStop}%
\bibitem [{ati()}]{atima}%
  \BibitemOpen
  \href@noop {} {}\bibinfo {note}
  {\lowercase{h}ttps://web-docs.gsi.de/$\sim$weick/atima/}\BibitemShut
  {NoStop}%
\bibitem [{\citenamefont {Jayatissa}\ \emph {et~al.}(2022)\citenamefont
  {Jayatissa}, \citenamefont {Avila}, \citenamefont {Rehm}, \citenamefont
  {Talwar}, \citenamefont {Mohr}, \citenamefont {Auranen}, \citenamefont
  {Chen}, \citenamefont {Gorelov}, \citenamefont {Hoffman}, \citenamefont
  {Jiang}, \citenamefont {Kay}, \citenamefont {Kuvin},\ and\ \citenamefont
  {Santiago-Gonzalez}}]{Jayatissa2022}%
  \BibitemOpen
  \bibfield  {author} {\bibinfo {author} {\bibfnamefont {H.}~\bibnamefont
  {Jayatissa}}, \bibinfo {author} {\bibfnamefont {M.~L.}\ \bibnamefont
  {Avila}}, \bibinfo {author} {\bibfnamefont {K.~E.}\ \bibnamefont {Rehm}},
  \bibinfo {author} {\bibfnamefont {R.}~\bibnamefont {Talwar}}, \bibinfo
  {author} {\bibfnamefont {P.}~\bibnamefont {Mohr}}, \bibinfo {author}
  {\bibfnamefont {K.}~\bibnamefont {Auranen}}, \bibinfo {author} {\bibfnamefont
  {J.}~\bibnamefont {Chen}}, \bibinfo {author} {\bibfnamefont {D.~A.}\
  \bibnamefont {Gorelov}}, \bibinfo {author} {\bibfnamefont {C.~R.}\
  \bibnamefont {Hoffman}}, \bibinfo {author} {\bibfnamefont {C.~L.}\
  \bibnamefont {Jiang}}, \bibinfo {author} {\bibfnamefont {B.~P.}\ \bibnamefont
  {Kay}}, \bibinfo {author} {\bibfnamefont {S.~A.}\ \bibnamefont {Kuvin}},\
  and\ \bibinfo {author} {\bibfnamefont {D.}~\bibnamefont
  {Santiago-Gonzalez}},\ }\href {https://doi.org/10.1103/PhysRevC.105.L042802}
  {\bibfield  {journal} {\bibinfo  {journal} {\prc}\ }\textbf {\bibinfo
  {volume} {105}},\ \bibinfo {eid} {L042802} (\bibinfo {year}
  {2022})}\BibitemShut {NoStop}%
\bibitem [{\citenamefont {Avila}\ \emph {et~al.}(2017)\citenamefont {Avila},
  \citenamefont {Rehm}, \citenamefont {Almaraz-Calderon}, \citenamefont
  {Ayangeakaa}, \citenamefont {Dickerson}, \citenamefont {Hoffman},
  \citenamefont {Jiang}, \citenamefont {Kay}, \citenamefont {Lai},
  \citenamefont {Nusair}, \citenamefont {Pardo}, \citenamefont
  {Santiago-Gonzalez}, \citenamefont {Talwar},\ and\ \citenamefont
  {Ugalde}}]{Avila2017}%
  \BibitemOpen
  \bibfield  {author} {\bibinfo {author} {\bibfnamefont {M.}~\bibnamefont
  {Avila}}, \bibinfo {author} {\bibfnamefont {K.}~\bibnamefont {Rehm}},
  \bibinfo {author} {\bibfnamefont {S.}~\bibnamefont {Almaraz-Calderon}},
  \bibinfo {author} {\bibfnamefont {A.}~\bibnamefont {Ayangeakaa}}, \bibinfo
  {author} {\bibfnamefont {C.}~\bibnamefont {Dickerson}}, \bibinfo {author}
  {\bibfnamefont {C.}~\bibnamefont {Hoffman}}, \bibinfo {author} {\bibfnamefont
  {C.}~\bibnamefont {Jiang}}, \bibinfo {author} {\bibfnamefont
  {B.}~\bibnamefont {Kay}}, \bibinfo {author} {\bibfnamefont {J.}~\bibnamefont
  {Lai}}, \bibinfo {author} {\bibfnamefont {O.}~\bibnamefont {Nusair}},
  \bibinfo {author} {\bibfnamefont {R.}~\bibnamefont {Pardo}}, \bibinfo
  {author} {\bibfnamefont {D.}~\bibnamefont {Santiago-Gonzalez}}, \bibinfo
  {author} {\bibfnamefont {R.}~\bibnamefont {Talwar}},\ and\ \bibinfo {author}
  {\bibfnamefont {C.}~\bibnamefont {Ugalde}},\ }\href
  {https://doi.org/10.1016/j.nima.2017.03.060} {\bibfield  {journal} {\bibinfo
  {journal} {Nucl. Instrum. Methods Phys. Res., Sect. A}\ }\textbf {\bibinfo
  {volume} {859}},\ \bibinfo {pages} {63} (\bibinfo {year} {2017})}\BibitemShut
  {NoStop}%
\bibitem [{\citenamefont {Ong}\ \emph {et~al.}(2022)\citenamefont {Ong},
  \citenamefont {Avila}, \citenamefont {Mohr}, \citenamefont {Rehm},
  \citenamefont {Santiago-Gonzalez}, \citenamefont {Chen}, \citenamefont
  {Hoffman}, \citenamefont {Meisel}, \citenamefont {Montes},\ and\
  \citenamefont {Pereira}}]{Ong2022}%
  \BibitemOpen
  \bibfield  {author} {\bibinfo {author} {\bibfnamefont {W.-J.}\ \bibnamefont
  {Ong}}, \bibinfo {author} {\bibfnamefont {M.~L.}\ \bibnamefont {Avila}},
  \bibinfo {author} {\bibfnamefont {P.}~\bibnamefont {Mohr}}, \bibinfo {author}
  {\bibfnamefont {K.~E.}\ \bibnamefont {Rehm}}, \bibinfo {author}
  {\bibfnamefont {D.}~\bibnamefont {Santiago-Gonzalez}}, \bibinfo {author}
  {\bibfnamefont {J.}~\bibnamefont {Chen}}, \bibinfo {author} {\bibfnamefont
  {C.~R.}\ \bibnamefont {Hoffman}}, \bibinfo {author} {\bibfnamefont
  {Z.}~\bibnamefont {Meisel}}, \bibinfo {author} {\bibfnamefont
  {F.}~\bibnamefont {Montes}},\ and\ \bibinfo {author} {\bibfnamefont
  {J.}~\bibnamefont {Pereira}},\ }\href
  {https://doi.org/10.1103/PhysRevC.105.055803} {\bibfield  {journal} {\bibinfo
   {journal} {\prc}\ }\textbf {\bibinfo {volume} {105}},\ \bibinfo {eid}
  {055803} (\bibinfo {year} {2022})}\BibitemShut {NoStop}%
\bibitem [{\citenamefont {Talwar}\ \emph {et~al.}(2018)\citenamefont {Talwar},
  \citenamefont {Bojazi}, \citenamefont {Mohr}, \citenamefont {Auranen},
  \citenamefont {Avila}, \citenamefont {Ayangeakaa}, \citenamefont {Harker},
  \citenamefont {Hoffman}, \citenamefont {Jiang}, \citenamefont {Kuvin},
  \citenamefont {Meyer}, \citenamefont {Rehm}, \citenamefont
  {Santiago-Gonzalez}, \citenamefont {Sethi}, \citenamefont {Ugalde},\ and\
  \citenamefont {Winkelbauer}}]{Talwar2018}%
  \BibitemOpen
  \bibfield  {author} {\bibinfo {author} {\bibfnamefont {R.}~\bibnamefont
  {Talwar}}, \bibinfo {author} {\bibfnamefont {M.~J.}\ \bibnamefont {Bojazi}},
  \bibinfo {author} {\bibfnamefont {P.}~\bibnamefont {Mohr}}, \bibinfo {author}
  {\bibfnamefont {K.}~\bibnamefont {Auranen}}, \bibinfo {author} {\bibfnamefont
  {M.~L.}\ \bibnamefont {Avila}}, \bibinfo {author} {\bibfnamefont {A.~D.}\
  \bibnamefont {Ayangeakaa}}, \bibinfo {author} {\bibfnamefont
  {J.}~\bibnamefont {Harker}}, \bibinfo {author} {\bibfnamefont {C.~R.}\
  \bibnamefont {Hoffman}}, \bibinfo {author} {\bibfnamefont {C.~L.}\
  \bibnamefont {Jiang}}, \bibinfo {author} {\bibfnamefont {S.~A.}\ \bibnamefont
  {Kuvin}}, \bibinfo {author} {\bibfnamefont {B.~S.}\ \bibnamefont {Meyer}},
  \bibinfo {author} {\bibfnamefont {K.~E.}\ \bibnamefont {Rehm}}, \bibinfo
  {author} {\bibfnamefont {D.}~\bibnamefont {Santiago-Gonzalez}}, \bibinfo
  {author} {\bibfnamefont {J.}~\bibnamefont {Sethi}}, \bibinfo {author}
  {\bibfnamefont {C.}~\bibnamefont {Ugalde}},\ and\ \bibinfo {author}
  {\bibfnamefont {J.~R.}\ \bibnamefont {Winkelbauer}},\ }\href
  {https://doi.org/10.1103/PhysRevC.97.055801} {\bibfield  {journal} {\bibinfo
  {journal} {\prc}\ }\textbf {\bibinfo {volume} {97}},\ \bibinfo {eid} {055801}
  (\bibinfo {year} {2018})}\BibitemShut {NoStop}%
\bibitem [{\citenamefont {Avila}\ \emph {et~al.}(2016)\citenamefont {Avila},
  \citenamefont {Rehm}, \citenamefont {Almaraz-Calderon}, \citenamefont
  {Ayangeakaa}, \citenamefont {Dickerson}, \citenamefont {Hoffman},
  \citenamefont {Jiang}, \citenamefont {Kay}, \citenamefont {Lai},
  \citenamefont {Nusair}, \citenamefont {Pardo}, \citenamefont
  {Santiago-Gonzalez}, \citenamefont {Talwar},\ and\ \citenamefont
  {Ugalde}}]{Avila2016}%
  \BibitemOpen
  \bibfield  {author} {\bibinfo {author} {\bibfnamefont {M.~L.}\ \bibnamefont
  {Avila}}, \bibinfo {author} {\bibfnamefont {K.~E.}\ \bibnamefont {Rehm}},
  \bibinfo {author} {\bibfnamefont {S.}~\bibnamefont {Almaraz-Calderon}},
  \bibinfo {author} {\bibfnamefont {A.~D.}\ \bibnamefont {Ayangeakaa}},
  \bibinfo {author} {\bibfnamefont {C.}~\bibnamefont {Dickerson}}, \bibinfo
  {author} {\bibfnamefont {C.~R.}\ \bibnamefont {Hoffman}}, \bibinfo {author}
  {\bibfnamefont {C.~L.}\ \bibnamefont {Jiang}}, \bibinfo {author}
  {\bibfnamefont {B.~P.}\ \bibnamefont {Kay}}, \bibinfo {author} {\bibfnamefont
  {J.}~\bibnamefont {Lai}}, \bibinfo {author} {\bibfnamefont {O.}~\bibnamefont
  {Nusair}}, \bibinfo {author} {\bibfnamefont {R.~C.}\ \bibnamefont {Pardo}},
  \bibinfo {author} {\bibfnamefont {D.}~\bibnamefont {Santiago-Gonzalez}},
  \bibinfo {author} {\bibfnamefont {R.}~\bibnamefont {Talwar}},\ and\ \bibinfo
  {author} {\bibfnamefont {C.}~\bibnamefont {Ugalde}},\ }\href
  {https://doi.org/10.1103/PhysRevC.94.065804} {\bibfield  {journal} {\bibinfo
  {journal} {\prc}\ }\textbf {\bibinfo {volume} {94}},\ \bibinfo {eid} {065804}
  (\bibinfo {year} {2016})}\BibitemShut {NoStop}%
\bibitem [{\citenamefont {McFadden}\ and\ \citenamefont
  {Satchler}(1966)}]{MCFADDEN1966}%
  \BibitemOpen
  \bibfield  {author} {\bibinfo {author} {\bibfnamefont {L.}~\bibnamefont
  {McFadden}}\ and\ \bibinfo {author} {\bibfnamefont {G.}~\bibnamefont
  {Satchler}},\ }\href
  {https://doi.org/https://doi.org/10.1016/0029-5582(66)90441-X} {\bibfield
  {journal} {\bibinfo  {journal} {Nucl. Phys.}\ }\textbf {\bibinfo {volume}
  {84}},\ \bibinfo {pages} {177} (\bibinfo {year} {1966})}\BibitemShut
  {NoStop}%
\bibitem [{\citenamefont {Mohr}(2015)}]{Mohr2015}%
  \BibitemOpen
  \bibfield  {author} {\bibinfo {author} {\bibfnamefont {P.}~\bibnamefont
  {Mohr}},\ }\href@noop {} {\bibfield  {journal} {\bibinfo  {journal} {Eur.
  Phys. J. A}\ }\textbf {\bibinfo {volume} {51}},\ \bibinfo {pages} {56}
  (\bibinfo {year} {2015})}\BibitemShut {NoStop}%
\bibitem [{\citenamefont {Haas}\ and\ \citenamefont {Bair}(1973)}]{Haas1973}%
  \BibitemOpen
  \bibfield  {author} {\bibinfo {author} {\bibfnamefont {F.~X.}\ \bibnamefont
  {Haas}}\ and\ \bibinfo {author} {\bibfnamefont {J.~K.}\ \bibnamefont
  {Bair}},\ }\href {https://doi.org/10.1103/PhysRevC.7.2432} {\bibfield
  {journal} {\bibinfo  {journal} {Phys. Rev. C}\ }\textbf {\bibinfo {volume}
  {7}},\ \bibinfo {pages} {2432} (\bibinfo {year} {1973})}\BibitemShut
  {NoStop}%
\bibitem [{\citenamefont {{Drotleff}}\ \emph {et~al.}(1993)\citenamefont
  {{Drotleff}}, \citenamefont {{Denker}}, \citenamefont {{Knee}}, \citenamefont
  {{Soine}}, \citenamefont {{Wolf}}, \citenamefont {{Hammer}}, \citenamefont
  {{Greife}}, \citenamefont {{Rolfs}},\ and\ \citenamefont
  {{Trautvetter}}}]{Drotleff1993}%
  \BibitemOpen
  \bibfield  {author} {\bibinfo {author} {\bibfnamefont {H.~W.}\ \bibnamefont
  {{Drotleff}}}, \bibinfo {author} {\bibfnamefont {A.}~\bibnamefont
  {{Denker}}}, \bibinfo {author} {\bibfnamefont {H.}~\bibnamefont {{Knee}}},
  \bibinfo {author} {\bibfnamefont {M.}~\bibnamefont {{Soine}}}, \bibinfo
  {author} {\bibfnamefont {G.}~\bibnamefont {{Wolf}}}, \bibinfo {author}
  {\bibfnamefont {J.~W.}\ \bibnamefont {{Hammer}}}, \bibinfo {author}
  {\bibfnamefont {U.}~\bibnamefont {{Greife}}}, \bibinfo {author}
  {\bibfnamefont {C.}~\bibnamefont {{Rolfs}}},\ and\ \bibinfo {author}
  {\bibfnamefont {H.~P.}\ \bibnamefont {{Trautvetter}}},\ }\href
  {https://doi.org/10.1086/173119} {\bibfield  {journal} {\bibinfo  {journal}
  {Astrophys. J.}\ }\textbf {\bibinfo {volume} {414}},\ \bibinfo {pages} {735}
  (\bibinfo {year} {1993})}\BibitemShut {NoStop}%
\bibitem [{\citenamefont {Matic}\ \emph {et~al.}(2010)\citenamefont {Matic},
  \citenamefont {van~den Berg}, \citenamefont {Harakeh}, \citenamefont
  {W\"ortche}, \citenamefont {Berg}, \citenamefont {Couder}, \citenamefont
  {G\"orres}, \citenamefont {LeBlanc}, \citenamefont {O'Brien}, \citenamefont
  {Wiescher}, \citenamefont {Fujita}, \citenamefont {Hatanaka}, \citenamefont
  {Sakemi}, \citenamefont {Shimizu}, \citenamefont {Tameshige}, \citenamefont
  {Tamii}, \citenamefont {Yosoi}, \citenamefont {Adachi}, \citenamefont
  {Fujita}, \citenamefont {Shimbara}, \citenamefont {Fujita}, \citenamefont
  {Wakasa}, \citenamefont {Brown},\ and\ \citenamefont {Schatz}}]{Matic2010}%
  \BibitemOpen
  \bibfield  {author} {\bibinfo {author} {\bibfnamefont {A.}~\bibnamefont
  {Matic}}, \bibinfo {author} {\bibfnamefont {A.~M.}\ \bibnamefont {van~den
  Berg}}, \bibinfo {author} {\bibfnamefont {M.~N.}\ \bibnamefont {Harakeh}},
  \bibinfo {author} {\bibfnamefont {H.~J.}\ \bibnamefont {W\"ortche}}, \bibinfo
  {author} {\bibfnamefont {G.~P.~A.}\ \bibnamefont {Berg}}, \bibinfo {author}
  {\bibfnamefont {M.}~\bibnamefont {Couder}}, \bibinfo {author} {\bibfnamefont
  {J.}~\bibnamefont {G\"orres}}, \bibinfo {author} {\bibfnamefont
  {P.}~\bibnamefont {LeBlanc}}, \bibinfo {author} {\bibfnamefont
  {S.}~\bibnamefont {O'Brien}}, \bibinfo {author} {\bibfnamefont
  {M.}~\bibnamefont {Wiescher}}, \bibinfo {author} {\bibfnamefont
  {K.}~\bibnamefont {Fujita}}, \bibinfo {author} {\bibfnamefont
  {K.}~\bibnamefont {Hatanaka}}, \bibinfo {author} {\bibfnamefont
  {Y.}~\bibnamefont {Sakemi}}, \bibinfo {author} {\bibfnamefont
  {Y.}~\bibnamefont {Shimizu}}, \bibinfo {author} {\bibfnamefont
  {Y.}~\bibnamefont {Tameshige}}, \bibinfo {author} {\bibfnamefont
  {A.}~\bibnamefont {Tamii}}, \bibinfo {author} {\bibfnamefont
  {M.}~\bibnamefont {Yosoi}}, \bibinfo {author} {\bibfnamefont
  {T.}~\bibnamefont {Adachi}}, \bibinfo {author} {\bibfnamefont
  {Y.}~\bibnamefont {Fujita}}, \bibinfo {author} {\bibfnamefont
  {Y.}~\bibnamefont {Shimbara}}, \bibinfo {author} {\bibfnamefont
  {H.}~\bibnamefont {Fujita}}, \bibinfo {author} {\bibfnamefont
  {T.}~\bibnamefont {Wakasa}}, \bibinfo {author} {\bibfnamefont {B.~A.}\
  \bibnamefont {Brown}},\ and\ \bibinfo {author} {\bibfnamefont
  {H.}~\bibnamefont {Schatz}},\ }\href
  {https://doi.org/10.1103/PhysRevC.82.025807} {\bibfield  {journal} {\bibinfo
  {journal} {Phys. Rev. C}\ }\textbf {\bibinfo {volume} {82}},\ \bibinfo
  {pages} {025807} (\bibinfo {year} {2010})}\BibitemShut {NoStop}%
\bibitem [{\citenamefont {Seweryniak}\ \emph {et~al.}(2007)\citenamefont
  {Seweryniak}, \citenamefont {Woods}, \citenamefont {Carpenter}, \citenamefont
  {Davinson}, \citenamefont {Janssens}, \citenamefont {Jenkins}, \citenamefont
  {Lauritsen}, \citenamefont {Lister}, \citenamefont {Shergur}, \citenamefont
  {Sinha},\ and\ \citenamefont {Woehr}}]{Seweryniak2007}%
  \BibitemOpen
  \bibfield  {author} {\bibinfo {author} {\bibfnamefont {D.}~\bibnamefont
  {Seweryniak}}, \bibinfo {author} {\bibfnamefont {P.~J.}\ \bibnamefont
  {Woods}}, \bibinfo {author} {\bibfnamefont {M.~P.}\ \bibnamefont
  {Carpenter}}, \bibinfo {author} {\bibfnamefont {T.}~\bibnamefont {Davinson}},
  \bibinfo {author} {\bibfnamefont {R.~V.~F.}\ \bibnamefont {Janssens}},
  \bibinfo {author} {\bibfnamefont {D.~G.}\ \bibnamefont {Jenkins}}, \bibinfo
  {author} {\bibfnamefont {T.}~\bibnamefont {Lauritsen}}, \bibinfo {author}
  {\bibfnamefont {C.~J.}\ \bibnamefont {Lister}}, \bibinfo {author}
  {\bibfnamefont {J.}~\bibnamefont {Shergur}}, \bibinfo {author} {\bibfnamefont
  {S.}~\bibnamefont {Sinha}},\ and\ \bibinfo {author} {\bibfnamefont
  {A.}~\bibnamefont {Woehr}},\ }\href
  {https://doi.org/10.1103/PhysRevC.75.062801} {\bibfield  {journal} {\bibinfo
  {journal} {Phys. Rev. C}\ }\textbf {\bibinfo {volume} {75}},\ \bibinfo
  {pages} {062801} (\bibinfo {year} {2007})}\BibitemShut {NoStop}%
\bibitem [{\citenamefont {Rauscher}()}]{Rauscher}%
  \BibitemOpen
  \bibfield  {author} {\bibinfo {author} {\bibfnamefont {T.}~\bibnamefont
  {Rauscher}},\ }\href@noop {} {\bibinfo {title} {Private comm.}}\BibitemShut
  {Stop}%
\bibitem [{REA()}]{REACLIB}%
  \BibitemOpen
  \href@noop {} {}\bibinfo {note} {JINA Reaclib Database
  \lowercase{h}ttps://reaclib.jinaweb.org}\BibitemShut {NoStop}%
\bibitem [{\citenamefont {{Merz}}\ and\ \citenamefont
  {{Meisel}}(2021)}]{Merz2021}%
  \BibitemOpen
  \bibfield  {author} {\bibinfo {author} {\bibfnamefont {G.}~\bibnamefont
  {{Merz}}}\ and\ \bibinfo {author} {\bibfnamefont {Z.}~\bibnamefont
  {{Meisel}}},\ }\href {https://doi.org/10.1093/mnras/staa3414} {\bibfield
  {journal} {\bibinfo  {journal} {Mon. Not. R. Astron. Soc.}\ }\textbf
  {\bibinfo {volume} {500}},\ \bibinfo {pages} {2958} (\bibinfo {year}
  {2021})}\BibitemShut {NoStop}%
\bibitem [{\citenamefont {{Ubertini}}\ \emph {et~al.}(1999)\citenamefont
  {{Ubertini}}, \citenamefont {{Bazzano}}, \citenamefont {{Cocchi}},
  \citenamefont {{Natalucci}}, \citenamefont {{Heise}}, \citenamefont
  {{Muller}},\ and\ \citenamefont {{in 't Zand}}}]{Ubertini1999}%
  \BibitemOpen
  \bibfield  {author} {\bibinfo {author} {\bibfnamefont {P.}~\bibnamefont
  {{Ubertini}}}, \bibinfo {author} {\bibfnamefont {A.}~\bibnamefont
  {{Bazzano}}}, \bibinfo {author} {\bibfnamefont {M.}~\bibnamefont {{Cocchi}}},
  \bibinfo {author} {\bibfnamefont {L.}~\bibnamefont {{Natalucci}}}, \bibinfo
  {author} {\bibfnamefont {J.}~\bibnamefont {{Heise}}}, \bibinfo {author}
  {\bibfnamefont {J.~M.}\ \bibnamefont {{Muller}}},\ and\ \bibinfo {author}
  {\bibfnamefont {J.~J.~M.}\ \bibnamefont {{in 't Zand}}},\ }\href
  {https://doi.org/10.1086/311933} {\bibfield  {journal} {\bibinfo  {journal}
  {Astrophys. J. Lett.}\ }\textbf {\bibinfo {volume} {514}},\ \bibinfo {pages}
  {L27} (\bibinfo {year} {1999})}\BibitemShut {NoStop}%
\end{thebibliography}%

\end{document}